\documentclass[aps,prl,preprintnumbers,superscriptaddress,nofootinbib,floatfix,10pt]{revtex4-1}
\usepackage{amsfonts,amssymb,stmaryrd,latexsym,amsmath,braket}
\usepackage{graphicx,subfigure}
\usepackage[export]{adjustbox}
\usepackage{times}
\usepackage{bbm}
\usepackage{bm}
\usepackage{mathrsfs}
\usepackage{slashed}
\usepackage[version=3]{mhchem}
\usepackage{braket}
\usepackage{hyperref}
\usepackage{verbatim}
\usepackage[utf8]{inputenc}

\usepackage{graphicx,subfigure}
\usepackage{times}
\usepackage{slashed}
\usepackage{braket}
\usepackage{verbatim}
\usepackage{multirow}
\usepackage[utf8]{inputenc}
\usepackage{array}
\newcommand{\PreserveBackslash}[1]{\let\temp=\\#1\let\\=\temp}
\newcolumntype{C}[1]{>{\PreserveBackslash\centering}p{#1}}
\newcolumntype{R}[1]{>{\PreserveBackslash\raggedleft}p{#1}}
\newcolumntype{L}[1]{>{\PreserveBackslash\raggedright}p{#1}}

\makeatletter
\let\orgdescriptionlabel\descriptionlabel
\renewcommand*{\descriptionlabel}[1]{%
	\let\orglabel\label
	\let\label\@gobble
	\phantomsection
	\edef\@currentlabel{#1}%
	\let\label\orglabel
	\orgdescriptionlabel{#1}%
}
\makeatother

\usepackage{color}

\definecolor{DarkMidnightBlue}{rgb}{0.0, 0.04, 0.14}

\hypersetup{hidelinks,
	breaklinks=true,
	colorlinks=true,
	linkcolor=blue, % default= red
	citecolor=blue, % default= green
	urlcolor=DarkMidnightBlue,
	bookmarksopen=false,    
	pdftitle={Title},
	pdfauthor={Author}}

\def\affHISKP{\affiliation{Helmholtz-Institut f\"ur Strahlen- und Kernphysik~(Theorie) and Bethe Center for Theoretical Physics, Universit\"at Bonn, D-53115 Bonn, Germany}}

\def\affGIBTU{\affiliation{Faculty of Natural Sciences and Engineering, Gaziantep Islam Science and Technology University, Gaziantep 27010, Turkey}}

\def\affJUELICH{\affiliation{
		Institute~for~Advanced~Simulation, Institut~f\"{u}r~Kernphysik,~ and J\"{u}lich~Center~for~Hadron~Physics, Forschungszentrum~J\"{u}lich, D-52425~J\"{u}lich,~Germany}}

\def\affTSU{\affiliation{Tbilisi State University, 0186 Tbilisi, Georgia}}

\def\affMSU{\affiliation{Facility for Rare Isotope Beams and Department of Physics and Astronomy,
		Michigan State University, MI 48824, USA}}

\def\affSCNU{\affiliation{Guangdong Provincial Key Laboratory of Nuclear Science, Institute of Quantum Matter, South China Normal University, Guangzhou 510006, China}}

\def\affGSCAEP{\affiliation{Graduate School of China Academy of Engineering Physics, Beijing
		100193, China}}

\def\affSYSU{\affiliation{School of Physics, Sun Yat-Sen University, Guangzhou 510275, China}}

\def\affBOCHUM{
	\affiliation{Institut f\"ur Theoretische Physik II, Ruhr-Universit\"at Bochum,
		D-44870 Bochum, Germany}}

\def\affRISP{\affiliation{Rare Isotope Science Project, Institute for Basic Science, Daejeon 34000, Korea}}

\def\affCENS{\affiliation{Center for Exotic Nuclear Studies, Institute for Basic Science, Daejeon 34126, Korea}}

\def\affMiSU{\affiliation{Department of Physics $\&$ Astronomy and HPC$^2$ Center for Computational Sciences, 
		Mississippi State University, Mississippi State, Mississippi State 39762, USA}}

\def\affGS{\affiliation{cDRF/IRFU/DPhN/LENA and dEspace de Structure Nucleaire Th\'eorique, \\ CEA Paris-Saclay, B\^{a}t. 703, 91190 Saint-Aubin, France}}

\begin{document}
	
	\title{Wavefunction matching for solving quantum many-body problems}

	\author{Serdar Elhatisari}
	\affGIBTU
	\affHISKP
	
	\author{Lukas Bovermann}
	\affBOCHUM
	
	\author{Yuanzhuo~Ma}
	\affSCNU

	\author{Evgeny~Epelbaum}
	\affBOCHUM
	
	\author{Dillon~Frame}
	\affJUELICH
	
	\author{Fabian Hildenbrand}
	\affJUELICH
	
	\author{Myungkuk Kim}
	\affCENS
	
	\author{Youngman Kim}
	\affRISP
	
	\author{Hermann~Krebs}
	\affBOCHUM
	
	\author{Timo~A.~L{\"a}hde}
	\affJUELICH
	
	\author{Dean~Lee}
	%\email{leed@frib.msu.edu}
	\affMSU
	
	\author{Ning~Li}
	\affSYSU
	
	\author{Bing-Nan~Lu}
	\affGSCAEP

	\author{Ulf-G.~Meißner}
	%\email{meissner@hiskp.uni-bonn.de}
	\affHISKP
	\affJUELICH
	\affTSU
	
	\author{Gautam Rupak}
	\affMiSU

	\author{Shihang~Shen}
	\affJUELICH
	
	\author{Young-Ho Song}
	\affRISP

	\author{Gianluca~Stellin}
	\affGS

	\begin{abstract}
		{\it Ab initio} calculations play an essential role in our fundamental understanding of quantum many-body systems across many subfields, from strongly correlated fermions \cite{assaad2008world,Schollwock:2011,Orus:2014} to quantum chemistry \cite{dovesi2005ab,friesner2005ab,bartlett2007coupled} and from atomic and molecular systems \cite{aymar1996multichannel,stone2007atom,motta2018ab} to nuclear physics \cite{barrett2013ab,hagen2014coupled,carlson2015quantum,hergert2016medium,stroberg2021ab}.  One of the primary challenges is to perform accurate calculations for systems where the interactions may be complicated and difficult for the chosen computational method to handle.  Here we address the problem by introducing a new approach called \textit{wavefunction matching}.  Wavefunction matching transforms the interaction between particles so that the wavefunctions up to some finite range match that of an easily computable interaction. This allows for calculations of systems that would otherwise be impossible due to problems such as Monte Carlo sign cancellations. We apply the method to lattice Monte Carlo simulations \cite{Lee:2008fa,lahde2019nuclear} of light nuclei, medium-mass nuclei, neutron matter, and nuclear matter. We use high-fidelity chiral effective field theory interactions \cite{Epelbaum:2008ga,Machleidt:2011} and find good agreement with empirical data.  These results are accompanied by new insights on the nuclear interactions that may help to resolve long-standing challenges in accurately reproducing nuclear binding energies, charge radii, and nuclear matter saturation in {\it ab initio} calculations \cite{Ekstrom:2022yea,Machleidt:2023jws}.
	\end{abstract}
	
	\maketitle
	
	Quantum Monte Carlo simulations are a powerful and efficient method that can describe strong correlations in quantum many-body systems.  No assumptions about the nature of the system are necessary, and the computational effort grows only as a low power of the number of particles.  For many problems of interest, a simple Hamiltonian $H^S$ can be found that describes the energies and other observables of the many-body system in fair agreement with empirical data and is easily computable using Monte Carlo methods. On the other hand, realistic high-fidelity Hamiltonians usually suffer from severe sign problems with positive and negative contributions to the averages cancelling each other, so that Monte Carlo calculations become impractical. In this work, we solve this problem using a new approach called wave function matching.  While keeping the observable physics unchanged, wave function matching creates a new high-fidelity Hamiltonian $H'$ such that wave functions at short distances match that of a simple Hamiltonian $H^S$ which is easily computed. This allows for a rapidly converging expansion in powers of the difference $H'-H^S$.  Wave function matching can be used with any computational scheme.  In the following analysis, we focus on the case of quantum Monte Carlo simulations, where the method presents a promising and practical strategy for evading the sign problem in realistic calculations of nuclear quantum many-body systems. 
	
\begin{figure}[htb]
		\centering\includegraphics[height=6 cm,valign=t]{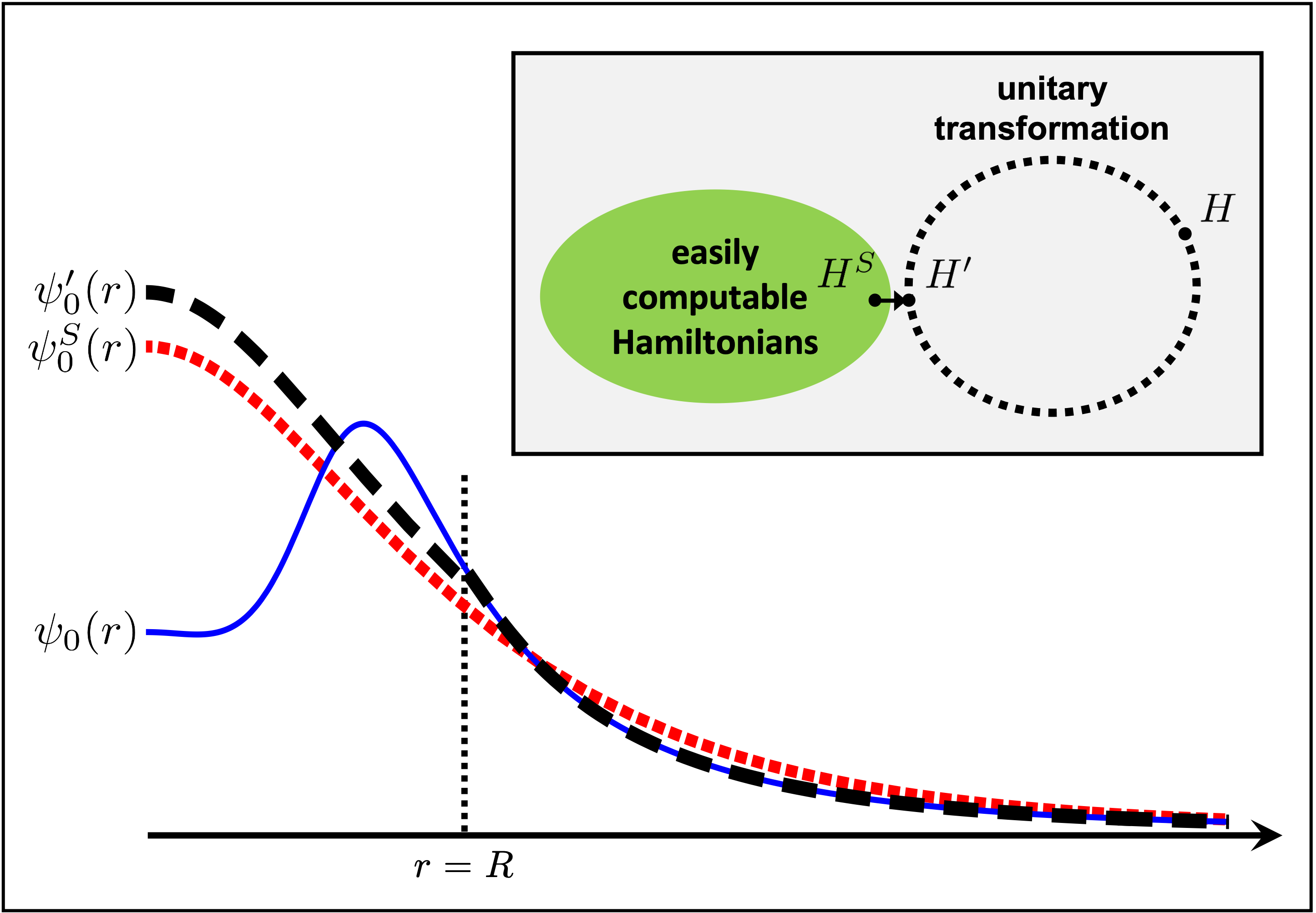}
		\hspace{0.5cm}
		\centering\includegraphics[height=6 cm,valign=t]{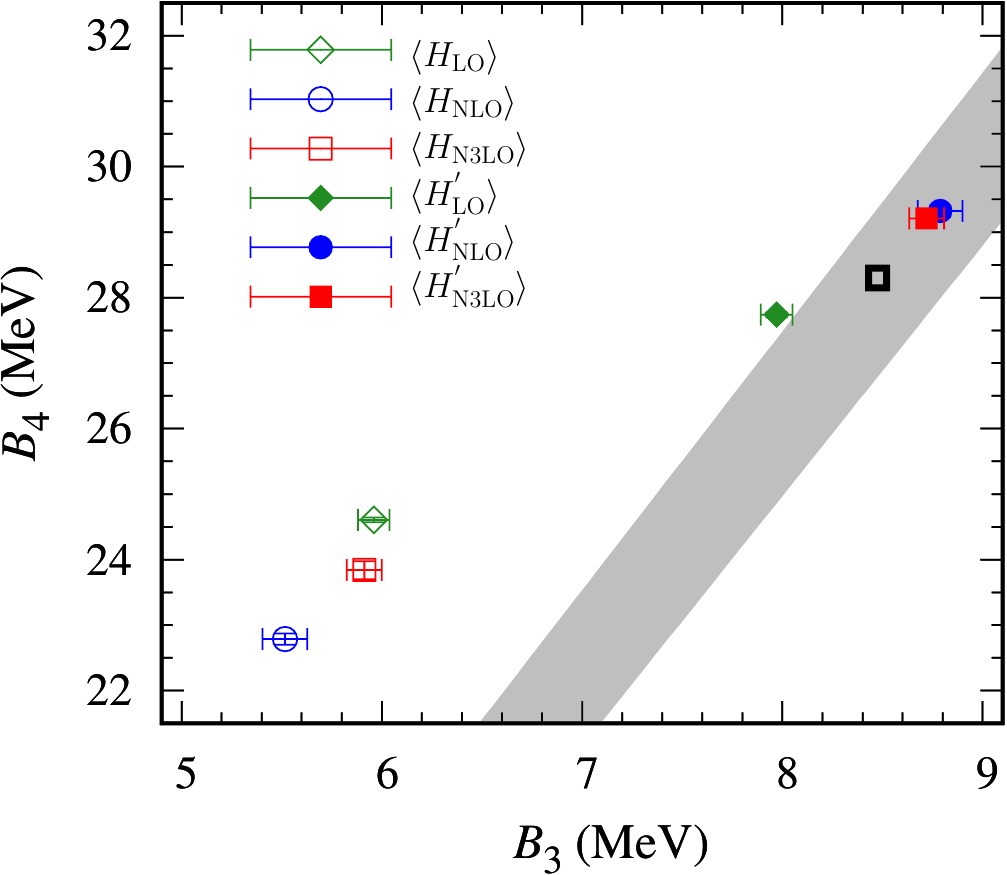}
		\caption{{\bf Left Panel: Pictoral representation of wave function matching.} The simple Hamiltonian $H^S$ is an easily computable Hamiltonian while the high-fidelity Hamiltonian $H$ is not.  A unitary transformation on the two-nucleon interaction with finite range $R$ is used to produce a new Hamiltonian $H'$ that is close to $H^S$. In each two-body channel, the ground state wave function of $H'$ matches the ground state wave function of $H$ for $r>R$ and is proportional to the ground state wave function of $H^S$ for $r<R$. 
			{\bf Right Panel: The Tjon band correlation between the binding energies of $^3$H and $^4$He.} The gray band is the predicted result from Ref.~\cite{Platter:2004zs}. The black open box shows the empirical point. The green-diamond, blue-circle, and read-square points show the results at LO, NLO, and N3LO in chiral effective field theory, respectively. The open points show the results from the first-order perturbative calculations using the Hamiltonian $H$, while the filled points are the results of the first-order perturbative calculations using the Hamiltonian $H'$.
			\label{fig:schematic}}
	\end{figure}

	A unitary transformation $U$ is a linear transformation that maps normalized orthogonal states to other normalized orthogonal states.
	Starting from a realistic high-fidelity Hamiltonian $H$, wave function matching defines a new Hamiltonian $H' = U^\dagger H U$, where $U^\dagger$ is the Hermitian conjugate of $U$.  In each two-body angular momentum channel, the unitary transformation $U$ is active only when the separation distance between two particles is less than some chosen distance $R$.  Let us write $\psi_0(r)$, $\psi'_0(r)$, and $\psi_0^S(r)$ for the ground state wave functions of $H$, $H'$, and the simple Hamiltonian $H^S$, respectively. The transformation $U$ is defined such that $\psi'_0(r)$ is proportional to $\psi_0^S(r)$ for $r < R$.  The simple Hamiltonian is chosen so that the constant of proportionality is close to $1$.  For $r > R$, however, $U$ is not active and so $\psi'_0(r)$ remains equal $\psi_0(r)$.  This is illustrated in the left panel of Fig.~\ref{fig:schematic}.

	Wave function matching will now be applied to {\it ab initio} Monte Carlo nuclear lattice simulations \cite{Lee:2008fa,Lahde:2019npb,Lu:2019nbg,Lu:2021tab,Shen:2022bak} using the framework of chiral effective field theory ($\chi$EFT) \cite{Epelbaum:2008ga,Machleidt:2016rvv}.  For our realistic Hamiltonian $H$, we use $\chi$EFT at next-to-next-to-next-to-leading order (N3LO) with lattice spacing $a=1.32$~fm.  For our simple Hamiltonian $H^S$, we employ a $\chi$EFT interaction at leading order.  Details of the interactions can be found in Methods. In the following, we use the term ``local'' for interactions that do not change the positions of particles while  ``nonlocal'' refers to interactions that do change the relative positions of particles.  The ``range'' of the interaction refers to the separation distance at which the interaction between particles becomes negligible.
	
	We calculate all quantities up to first order in perturbation theory, which corresponds to one power in the difference $H'-H^S$.  As a first test, we consider the energy of the deuteron, $^2$H.  The wave function matching calculation gives a binding energy of $2.02$~MeV, in comparison with $2.21$~MeV for the true binding energy of $H$ and $2.22$~MeV for the experimentally observed value. The residual error of $0.1$~MeV per nucleon is due to corrections beyond first order in powers of $H'-H^S$.  If one does not use wave function matching and instead performs the analogous calculation to first order in $H-H^S$, the result is a much less accurate binding energy of $0.68$~MeV. 
	
	As a second test of wave function matching, we calculate the binding energies of $^3$H and $^4$He. The Tjon band describes the universal correlations between the $^3$H and $^4$He binding energies \cite{Tjon:1975sme,Platter:2004zs}.  Provided that there are no long-range nonlocal interactions, any realistic two-nucleon interaction produces binding energies that lie on the Tjon band.  The inclusion of any short-range three-nucleon interaction also preserves this universal relation.  In Fig.~\ref{fig:schematic} we show wave function matching calculations using two-nucleon interactions only.  At leading order (LO) the calculated point falls outside the Tjon band since the Coulomb interaction is not included, while the next-to-leading order (NLO) and N3LO results lie squarely in the middle of the band. We are using a low-energy scheme where the two-nucleon interaction is the same at NLO and next-to-next-to-leading order (NNLO). \cite{Li:2018ymw}.  The empirical point is also shown in Fig.~\ref{fig:schematic}. The good agreement with the Tjon band suggests a residual error of $0.1$~MeV per nucleon or less.  This can be compared with the substantial deviation from the Tjon line if one does not use wave function matching and performs the analogous calculation to first order in $H-H^S$.
	
	Before proceeding to larger nuclei and many-body systems, we first comment on the current status of {\it ab initio} calculations of nuclear structure using $\chi$EFT.  There has been tremendous progress in the last few years towards producing accurate results for nuclear structure across much of the nuclear chart using a variety of different computational approaches \cite{Drischler:2017wtt,Lonardoni:2017hgs,Morris:2017vxi,Piarulli:2017dwd,Soma:2019bso,Gysbers:2019uyb,Maris:2020qne,Hebeler:2020ocj,Jiang:2020the,Wirth:2021pij}.  But there is also ample evidence that the calculations are sensitive to the manner in which the short-distance features of the interactions are regulated \cite{Stroberg:2016ung,Huther:2019ont,Hoppe:2019uyw,Nosyk:2021pxb}, a warning sign that systematic errors are not fully under control.  Furthermore, no {\it ab initio} calculations have been able to provide an accurate description of nuclear binding energies, charge radii, and the saturation properties of symmetric nuclear matter with equal numbers of protons and neutrons.  Our goal here is to identify the problem and point to a viable solution.
	
	The results in Ref.~\cite{Elhatisari:2016owd,Kanada-Enyo:2020zzf} showed that the range and locality of the nuclear interactions have a strong influence on nuclear binding.  The $^4$He nucleus, also called an $\alpha$ particle, is a spatially compact object whose radius is comparable to the range of the interactions between nucleons.  The central density of the $\alpha$ particle is about $50\%$ greater than the saturation density of nuclear matter.  As a result, the $\alpha\alpha$ interaction is highly sensitive to the range and locality of the nucleonic interactions.  These same arguments apply to other interactions involving $\alpha$ particles and nucleons, which we denote as $N$.  Using the formalism of cluster effective field theory (also called halo effective field theory) \cite{Bertulani:2002sz,Higa:2008dn,Rotureau:2012yu,Hammer:2017tjm} for $\alpha$ particles and nucleons, the two-cluster interactions are $\alpha\alpha$ and $\alpha N$, and the three-cluster interactions are $\alpha\alpha\alpha$, $\alpha\alpha N$, and $\alpha N N$, with $\alpha N N$ having two possible isospin channels.  In addition to the $\alpha\alpha$ interaction, there will be some dependence on short-distance physics arising from these five cluster interactions in their most attractive channels.  While cluster effective field theory is not well suited for describing the structure of heavy nuclei and nuclear matter, it does provides a useful framework for categorizing systematic errors in nuclear many-body systems due to short-distance features of the interactions.
	
\begin{figure}[tb]
		\centering\includegraphics[height=4.2 cm,valign=t]{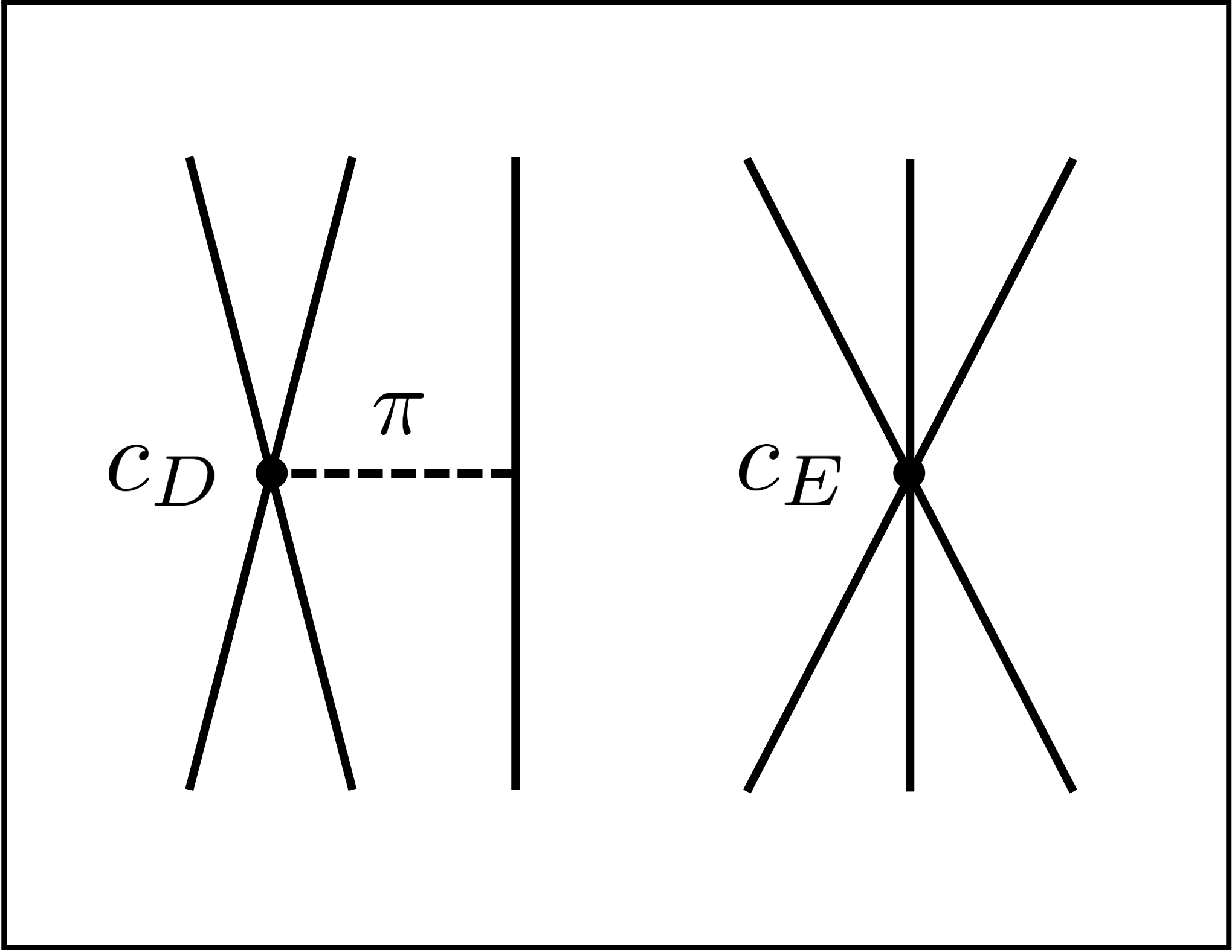}
		\hspace{0.5cm}
		\includegraphics[height=5.1 cm,valign=t]{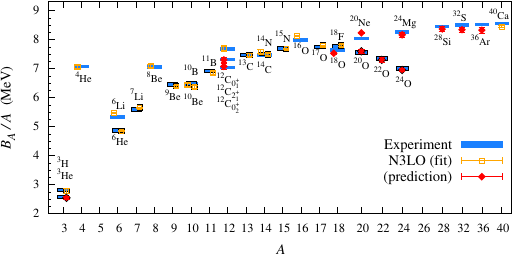}
		
		\caption{{\bf Left Panel:  Short-range three nucleon forces at NNLO.}  The first is the one-pion exchange term $c_D$ shown on the left.  The other is the purely short-range term $c_E$ shown on the right.  At order N3LO there are additional three-nucleon interactions associated with the exchange of two pions, as well as the corrections from the renormalization of the $c_D$ and $c_E$ terms at N3LO order.
			{\bf Right Panel: Results for nuclear binding energies using wave function matching.} Calculated ground state  and excited state energies of some selected nuclei with up to $A = 40$ at N3LO in chiral effective field theory and comparison with experimental data. The symbols with a black border indicate nuclei with unequal numbers of protons and neutrons.  The nuclei used in the fit of the higher-order three-nucleon interactions are labelled with open squares, while the other nuclei are predictions denoted with filled diamonds.
			\label{fig:threeN}}
	\end{figure}

	We will remove the unwanted dependence on short-distance features of the nucleonic interactions by fitting the binding energies of several light and medium-mass nuclei using three-nucleon interactions that go beyond N3LO order.  In $\chi$EFT, three nucleon forces first appear at order NNLO.  These include terms associated with the exchange of two pions and whose coefficients are determined from pion-nucleon scattering.  There are also two interactions with singular short-distance properties that must be regulated and the corresponding couplings fitted to empirical data.  As shown in Fig.~\ref{fig:threeN}, $c_D$ corresponds to the short-range interaction of two nucleons linked to a third nucleon through the exchange of a pion, and $c_E$ corresponds to the short-range interaction of all three nucleons.  At order N3LO, there are additional terms associated with the exchange of two pions as well as readjustments of the $c_D$ and $c_E$ coefficients at N3LO order~\cite{Ishikawa:2007zz,Bernard:2007sp,Bernard:2011zr}.  
	The higher-order three-nucleon interactions we consider are short-range modifications to the $c_D$ and $c_E$ terms that are designed to be strongly correlated with the cluster interactions described above.  In Methods we present the details of these new interactions along with a detailed uncertainty analysis. We find that with just one parameter, the root-mean-square-deviation (RMSD) for the energy per nucleon drops from about $2$~MeV down to $0.4$~MeV.  With the addition of five additional parameters, the RMSD per nucleon drops further to $0.1$~MeV.  These results are consistent with the hypothesis that the $\alpha\alpha$ interaction plays a key role in nuclear binding and that there are five additional cluster interactions that are sensitive to short distance physics.
	
	In the right panel of Fig.~\ref{fig:threeN}, we present the results for the nuclear binding energies using wave function matching.  We show ground state  and excited state energies of selected nuclei with up to $A = 40$ nucleons and comparison with experimental data. The symbols with a black border indicate nuclei with unequal numbers of protons and neutrons.  The nuclei used in the fit of the higher-order three-nucleon interactions are labelled with open squares, while the other nuclei are predictions denoted with filled diamonds. The one-standard-deviation error bars shown in Fig.~\ref{fig:threeN} represent uncertainties due to computational uncertainties from Monte Carlo errors, infinite volume extrapolation, and infinite projection time extrapolation.  We estimate the additional systematic errors due to truncation of the expansion of the $H'-H^S$ to be approximately $0.1$~MeV per nucleon.  The theoretical uncertainty due to the fitting of N3LO interactions have been empirically reduced by fitting the higher-order three-nucleon interactions to the binding energies of several nuclei.  We estimate the residual uncertainty due to the interactions to be less than $0.1$~MeV per nucleon.  
	
\begin{figure}[tb]
		\centering\includegraphics[height = 5.3 cm,valign=t]{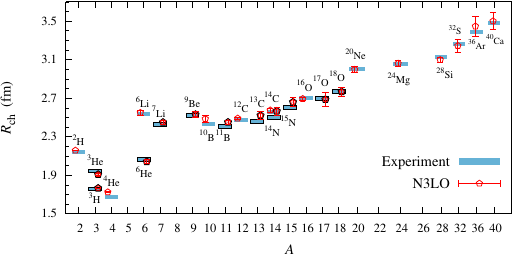}
		\hspace{0.5cm}
		\centering\includegraphics[height = 5.3 cm,valign=t]{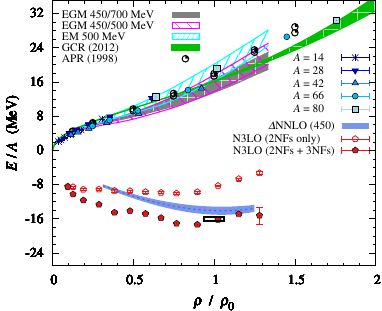}
		\caption{
			{\bf Left Panel: Predictions for charge radii of nuclei up to $A = 40$ at N3LO in chiral effective field theory and comparison with experimental data.} The symbols with a black border indicate nuclei with unequal numbers of protons and neutrons.
			{\bf Right Panel: Predictions for pure neutron matter energy per neutron and symmetric nuclear matter energy per nucleon as a function of density at N3LO in chiral effective field theory.} For the pure neutron matter energy we use the number of neutrons from $14$ to $80$ and various box sizes from $6.58$~fm to $13.2$~fm, and  for the symmetric nuclear matter energy we use the number of nucleons from $12$ to $160$ and a periodic box of length $9.21$~fm. For comparison we show the results from variational (APR)~\cite{Akmal:1998cf}, Auxiliary Field Diffusion MC calculations (GCR)~\cite{Gandolfi:2011xu}, calculated with N3LO/NNLO (two-nucleon/three-nucleon) chiral interactions (EM 500 MeV, EGM 450/500 MeV and EGM 450/700 MeV)~\cite{Tews:2012fj} and NNLO chiral interactions with explicit delta degrees of freedom ($\Delta$NNLO)~\cite{Ekstrom:2017koy}.
			\label{fig:ChargeRadii}}
	\end{figure}

	In the left panel of Fig.~\ref{fig:ChargeRadii}, we present the results for the charge radii of nuclei with up to $A = 40$ nucleons.  No charge radii data were used to fit any interaction parameters.  The one-standard-deviation error bars shown in Fig.~\ref{fig:ChargeRadii} represent computational uncertainties due to Monte Carlo errors, infinite volume extrapolation, and infinite projection time extrapolation.  The theoretical systematic uncertainties for the charge radii calculations are more difficult to predict without a comprehensive study of the dependence of charge radii upon the individual interactions.  However, the agreement with empirical results is quite good, with an RMSD of about $0.03$~fm. Note that the larger errors for the heaviest nuclei are statistical and can be decreased by utilizing greater computational resources.
	
	In the right panel of Fig.~\ref{fig:ChargeRadii}, we present lattice results for the energy per nucleon versus density for  pure neutron matter and symmetric nuclear matter.  None of the neutron matter and symmetric nuclear matter data were used to fit any interaction parameters. The density is expressed as a fraction of the saturation density for nuclear matter, $\rho_0 = 0.16$~fm$^{-3}$.  For the neutron matter calculations, we consider $14$ to $80$ neutrons in periodic box lengths ranging from $6.58$~fm to $13.2$~fm.  For the symmetric nuclear matter calculations, we use system sizes from $12$ to $160$ nucleons in a periodic box of length $9.21$~fm.  The comparisons with several other published work are shown and detailed in the figure caption.  We see that the neutron matter calculations are in good agreement with previous calculations, and the symmetric nuclear matter calculations pass through the empirical saturation point.  The one-standard-deviation error bars represent computational uncertainties due to Monte Carlo errors and infinite projection time extrapolation.  Uncertainties associated with extrapolation to large system size are not included here and will be explored in future work.  These lattice simulations of symmetric nuclear matter are qualitatively different from other theoretical calculations that assume a homogeneous phase.  The lattice simulations exhibit phase separation and cluster formation, just as in the real physical system.  Due to the finite number of nucleons in these calculations, some oscillations due to nuclear shell effects can be seen in the energy per nucleon.
	
	Another interesting feature of the lattice results is that symmetric nuclear matter without three-nucleon forces is underbound rather than overbound.  This is opposite to what is found in other calculations using renormalization group (RG) methods to ``soften'' the nucleonic interactions \cite{Bogner:2001gq,Bogner:2006pc,Bogner:2010}.  The term ``soft'' refers to the reduction of the strength of the interaction at large momentum values.  The difference arises from the fact that wave function matching produces a two-nucleon interaction that is more nonlocal and also less ``soft'' when compared with two-nucleon interactions produced using RG flow equations. 
	While there are some parallels in design, wave function matching is qualitatively different from RG methods.  Wave function matching implements a unitary transformation that has finite range, and so the transformation can be viewed as redefining the $\chi$EFT two-nucleon interaction.  One does not need to explicitly compute the many-body forces induced by the unitary transformation and needs only to determine the new coefficients of the short-range $\chi$EFT three-nucleon and higher-nucleon forces.  This has been demonstrated in a toy model~\cite{Bovermann:2022ohf}.  The approach used in wave function matching is similar to that of the Unitary Correlation Operator Method (UCOM) \cite{Feldmeier:1997zh,Neff:2002nu,Roth:2010bm}, though the operator-based unitary transformations in UCOM are significantly different in nature and purpose.
	
	In summary, we have presented a new approach for solving quantum many-body systems called wave function matching.  Wave function matching uses a transformation of the particle interactions to allow for calculations of systems that would otherwise be difficult or impossible.  We have applied the method to lattice Monte Carlo simulations of light nuclei, medium-mass nuclei, neutron matter, and nuclear matter at N3LO in $\chi$EFT and found good agreement with empirical data.  We have discussed the connection between systematic errors in nuclear many-body systems at short distances and interactions of $\alpha$ particles with themselves and nucleons.  These new developments may help resolve long-standing challenges in accurately reproducing nuclear binding energies, charge radii, and nuclear matter saturation in {\it ab initio} calculations.  While we have focused on Monte Carlo simulations for nuclear physics here, wave function matching can be used with any computational method applied to any quantum many-body system.  This also includes quantum computing algorithms where wave function matching can be used to reduce the number of quantum gates required.  All that is needed is a simple Hamiltonian $H^S$ that produces fair agreement with empirical data for the many-body system of interest and is easily computable using the method of choice.  Further details on the implementation of wave function matching are given in Methods.
	
	{\it We are grateful for discussions with members and partners of the Nuclear Lattice Effective Field Theory Collaboration (Joaquín Drut, Gustav Jansen, Stefan Krieg, Zhengxue Ren, Avik Sarkar, and Qian Wang) as well as Scott Bogner, Andreas Ekstr{\"o}m, Heiko Hergert, Morten Hjorth-Jensen, Daniel Phillips, Achim Schwenk, and Witek Nazarewicz.
		We acknowledge funding by  the Deutsche Forschungsgemeinschaft
		(DFG, German Research Foundation) and the NSFC through the funds provided  to  the  Sino-German
		Collaborative  Research  Center  TRR110  ``Symmetries  and  the  Emergence  of  Structure in  QCD''
		(DFG  Project  ID 196253076  -  TRR  110,  NSFC Grant  No.  12070131001),
		the Chinese Academy of Sciences (CAS) President's International Fellowship Initiative (PIFI)
		(Grant No. 2018DM0034), Volkswagen Stiftung  (Grant  No.  93562),  the European Research Council (ERC) under the
		European Union's Horizon 2020 research and innovation programme (ERC AdG EXOTIC, grant agreement No. 101018170, and ERC AdG NuclearTheory, grant agreement No. 426661267),  the Scientific and Technological Research Council of Turkey (TUBITAK project no. 120F341), the National Natural Science Foundation of China (Grants No. 12105106 and No. 12275259), U.S. National Science Foundation (PHY-1913620 and PHY-2209184),
		U.S. Department of Energy (DE-SC0013365 and DE-SC0021152), the Nuclear Computational Low-Energy Initiative (NUCLEI) SciDAC-4 project (DE-SC0018083), the Rare Isotope Science Project of the Institute for Basic Science funded by the Ministry of Science and ICT (MSICT) and by the National Research Foundation of Korea (2013M7A1A1075764), and the Institute for Basic Science (IBS-R031-D1-2022-a00).  Computational resources provided by the Gauss
		Centre for Supercomputing e.V. (www.gauss-centre.eu) for computing time on the GCS Supercomputer JUWELS at J{\"u}lich
		Supercomputing Centre (JSC) and special GPU time allocated on JURECA-DC as well as
		the Oak Ridge Leadership Computing Facility through the INCITE award
		``Ab-initio nuclear structure and nuclear reactions'', and partially provided by TUBITAK ULAKBIM High Performance and Grid Computing Center (TRUBA resources).
		vuComputational resources were also partly provided by the National Supercomputing Center of Korea with supercomputing resources including technical support ( KSC-2021-CRE-0429 ,  KSC-2022-CHA-0003). The work of Y.H.S. and Y.K. was supported by the Rare Isotope Science Project of Institute for Basic Science, funded by Ministry of Science and ICT (MSICT) and by National Research Foundation of Korea (2013M7A1A1075764). M.K. is supported by the Institute for Basic Science (IBS-R031-D1-2022-a00)}.
	\clearpage
	
	\section{Methods}

	\subsection{Hamiltonian Translators}\label{S-2.0}
	
	Before describing how wave function matching is implemented in practice, we first discuss a class of transformations called Hamiltonian translators.  Let $H_A$ and $H_B$ be two Hamiltonians acting on the same linear space.  Suppose that $U_{AB}$ is a unitary transformation mapping eigenvectors of $H_B$ to the eigenvectors of $H_A$.  We then call $U_{AB}$ a  Hamiltonian translator from $H_B$ to $H_A$. Clearly, $U_{BA}=U_{AB}^\dagger$ is then a Hamiltonian translator from $H_A$ to $H_B$.  We note the curious fact that $H'_A = U_{BA} H_A U_{AB}$ is a Hamiltonian with energy eigenvalues the same as $H_A$ but with eigenvectors corresponding to $H_B$.  Likewise, $H'_B = U_{AB} H_B U_{BA}$ is a Hamiltonian with energy eigenvalues the same as $H_B$ but with the eigenvectors corresponding to $H_A$.  
	
	Since $H'_A$ and $H_B$ share the same eigenvectors, $H'_A$ and $H_B$ commute with each other. In order to compute any energy eigenvalue of $H'_A$, it suffices to prepare the corresponding eigenvector of $H_B$ and compute the energy expectation value of $H'_A$.  We can express these facts using the language of perturbation theory.  If we write $H'_A = H_B + (H'_A - H_B)$, then the zeroth-order expansion of the eigenvectors is exact and the first-order expansion of the energies is exact.
	
	We can construct Hamiltonian translators using quantum adiabatic evolution \cite{Farhi:2000a}.  Let $f(t)$ be a smooth function such that $f(0)=0$ and $f(1)=1$.  Then for any $T > 0$, we can define the time-dependent Hamiltonian $H_T(t) = f(\tfrac{t}{T}) H_A + [1-f(\tfrac{t}{T})] H_B$.  We also define the unitary transformation
	\begin{equation}
	U_T = \overleftarrow{T} \exp \left[-i \int_{0}^{T} H_T(t)dt \right],
	\end{equation}
	where $\overleftarrow{T}$ is the time ordering symbol placing operators at later times on the left.  In the limit of large $T$, $U_T$ is a Hamiltonian translator from $H_B$ to $H_A$.  In the limit of large $T$, $U_T$ maps every eigenvector of $H_B$ to an eigenvector of $H_A$.  Within each symmetry subspace that is invariant under both Hamiltonians $H_A$ and $H_B$, the mapping $U_T$ preserves the ordering of energy eigenvalues.
	
	\subsection{Wave Function Matching}
	
	As discussed in the main text, we let $H^S$ be a simple Hamiltonian that is easily computable and $H$ be a Hamiltonian with realistic interactions.  Wave function matching can be viewed as an approximate Hamiltonian translator from $H^S$ to $H$.  It is only an approximate translator because the unitary transformation $U$ will be restricted to the space of two nucleons up to some maximum separation distance $R$ and will only map the lowest eigenvector of $H^S$ to the lowest eigenvector of $H$ in each scattering channel.
	
	In the following, we restrict our focus to the space of two nucleons in some angular-momentum scattering channel.  We impose hard wall boundary conditions at some very large separation distance $R_{\rm wall}$.  The energy eigenstates of $H$ for our chosen two-nucleon scattering channel will be denoted $\ket{\psi_n}$ for $n=0,1,\cdots$, and $E_n$ will be the corresponding energy eigenvalues.  We let the corresponding energy eigenstates for $H^S$ be denoted $\ket{\psi^S_n}$.  We now define a short-distance projection operator $P_R$ that projects out the portion of the two-nucleon state with separation distance less than or equal to $R$. We let $\ket{m}$ for $m=1,\cdots, m_R$ be an orthogonal basis spanning the set of two-nucleon channel states at short distances so that $P_R = \sum_{m=1}^{m_R} \ket{m}\bra{m}$.  In the lattice calculations presented here, we take $\ket{m}$ to be radial position eigenstates, sorted according to increasing radial distance.  See, for example, Ref.~\cite{Elhatisari:2015iga} for a discussion of radial position eigenstates on the lattice.  For continuum space calculations, it is more convenient to work with smooth basis functions.  For example, one can take $\ket{m}$ to be energy eigenstates sorted in order of increasing energy for a steep harmonic oscillator potential.  The projection operator $P_R = \sum_{m=1}^{m_{R}} \ket{m}\bra{m}$ does not have strictly finite range, but nevertheless diminishes very rapidly as a function of distance.  We take $R$ to be the distance at which the interaction is negligible.
	
	Let $\ket{\psi_0}_R$ and $\ket{\psi^S_0}_R$ be the normalized short-distance portions of the ground states of $H$ and $H^S$ respectively, 
	\begin{equation}
	\ket{\psi_0}_R = \frac{P_R\ket{\psi_0}}{\lVert P_R\ket{\psi_0} \rVert},   \; \;    \ket{\psi^S_0}_R = \frac{P_R\ket{\psi^S_0}}{\lVert P_R\ket{\psi^S_0} \rVert}.
	\end{equation}
	Let us define a unitary transformation $U$ such that $\ket{\psi^S_0}_R$ is mapped to $\ket{\psi_0}_R$.  There are several ways to define the remaining properties of $U$.  In this work, we use Gram-Schmidt orthogonalization to define an ordered sequence of orthonormal basis states, $\{\ket{\psi_0}_R, \ket{1}_\perp, \cdots, \ket{m_R-1}_\perp\}$ and, similarly, $\{\ket{\psi^S_0}_R, \ket{1}^S_\perp, \cdots, \ket{m_R-1}^S_\perp\}$.  We require that $U$ maps each basis vector $\ket{j}^S_\perp$ to the corresponding basis vector $\ket{j}_\perp$ for each $j = 1, \cdots, m_R-1$.
	
	We proceed by defining the transformed Hamiltonian $H' = U^\dagger H U$.  Let $\ket{\psi'_0}$ be the ground state of $H'$ and let 
	\begin{equation}
	\ket{\psi'_0}_R = \frac{P_R\ket{\psi'_0}}{\lVert P_R\ket{\psi'_0} \rVert}.
	\end{equation}
	We note that $\ket{\psi'_0} = U^\dagger\ket{\psi_0}$, ignoring any irrelevant overall phases.
	Since $U$ is negligible at distances greater than $R$, $\ket{\psi'_0}$ must equal $\ket{\psi_0}$ at distances greater than $R$.  At distances less than $R$, $\ket{\psi'_0}_R$ must equal $\ket{\psi^S_0}_R$.  Hence the short distance behavior of $\ket{\psi'_0}$ must be proportional to $\ket{\psi^S_0}$, with constant of proportionality $\kappa$.  If $\ket{\psi_0}$ and $\ket{\psi^S_0}$ have approximately equal normalizations at distances greater than $R$, then $\kappa$ will be numerically close to $1$, ignoring any irrelevant overall phases.
	
	We note that wave function matching introduces some nonlocality to the two-nucleon interactions in $H'$.  In principle, the process of wave function matching could increase the range of the interactions up to a distance scale equal to $R$. In practice, however, the range of interactions in $H'$ is determined by the radial distances for which the difference between the wave functions $\ket{\psi_0}$ and $\ket{\psi^S_0}$ is significant.  For the lattice calculations presented here we take $R$ to be $3.7$~fm, while the difference between $\ket{\psi_0}$ and $\ket{\psi^S_0}$ is appreciable at distances less than $1.9$~fm.
	
	\subsection{Lattice Operators}\label{S-3.0}
	
	\subsubsection{Simple Hamiltonian}
	
	We now present the details of our simple Hamiltonian $H^S$. We construct the Hamiltonian using a $\chi$EFT interaction at leading order,
	\begin{comment}
	\begin{equation}
	H^S=K+\frac{c_{{\rm 2N}}}{2}\sum_{\vec{n},\vec{n}^{\prime}}:
	\tilde{\rho}(\vec{n})
	f_{\rm s_{L}}^{(1)}(\vec{n}-\vec{n}^{\prime})
	\tilde{\rho}(\vec{n}^{\prime})
	(\vec{n}): + V_{\rm OPE}^{\Lambda_{\pi}} ,
	\label{eq:Hsimple}
	\end{equation}
	\end{comment}
	\begin{equation}
	H^S=K+\frac{c_{{\rm SU(4)}}}{2}\sum_{\vec{n}}:
	\left[
	\tilde{\rho}^{(1)}(\vec{n})
	\right]^2
	:  
	+\frac{c_{I}}{2}\sum_{I,\vec{n}}:
	\left[
	\tilde{\rho}^{(1)}_{I}(\vec{n})
	\right]^2
	: + 
	V_{\rm OPE}^{\Lambda_{\pi}} ,
	\label{eq:Hsimple2}
	\end{equation}
	where $K$ is the kinetic energy term with nucleon mass $m=938.92$ MeV, the $::$ symbol indicates normal ordering,
	and $\tilde{\rho}^{(d)}$ and $\tilde{\rho}^{(d)}_{I}$ are density operators that are smeared both locally and non-locally,
	\begin{equation}
	\tilde{\rho}^{(d)}(\vec{n}) = \sum_{i,j=0,1} 
	\tilde{a}^{\dagger}_{i,j}(\vec{n}) \, \tilde{a}^{\,}_{i,j}(\vec{n})
	+
	s_{\rm L}
	\sum_{|\vec{n}-\vec{n}^{\prime}|^2 = 1}^d 
	\,
	\sum_{i,j=0,1} 
	\tilde{a}^{\dagger}_{i,j}(\vec{n}^{\prime}) \, \tilde{a}^{\,}_{i,j}(\vec{n}^{\prime})
	\,,
	\end{equation}
	\begin{equation}
	\tilde{\rho}^{(d)}_{I}(\vec{n}) = \sum_{i,j,j^{\prime}=0,1} 
	\tilde{a}^{\dagger}_{i,j}(\vec{n}) \,\left[\tau_{I}\right]_{j,j^{\prime}} \, \tilde{a}^{\,}_{i,j^{\prime}}(\vec{n})
	+
	s_{\rm L}
	\sum_{|\vec{n}-\vec{n}^{\prime}|^2 = 1}^d 
	\,
	\sum_{i,j,j^{\prime}=0,1} 
	\tilde{a}^{\dagger}_{i,j}(\vec{n}^{\prime}) \,\left[\tau_{I}\right]_{j,j^{\prime}} \, \tilde{a}^{\,}_{i,j^{\prime}}(\vec{n}^{\prime})
	.
	\end{equation}
	The smeared annihilation and creation operators, $\tilde{a}$ and $\tilde{a}^{\dagger}$, have with spin $i = 0, 1$ (up, down) and isospin $j = 0, 1$ (proton, neutron) indices,
	\begin{equation}
	\tilde{a}_{i,j}(\vec{n})=a_{i,j}(\vec{n})+s_{\rm NL}\sum_{|\vec{n}^{\prime}-\vec{n}|=1}a_{i,j}(\vec{n}^{\prime}).
	\end{equation}
	Through our calculations we use local smearing parameter $s_{\rm L}=0.07$ and nonlocal smearing parameter $s_{\rm NL} = 0.5$. In addition to the short-range SU(4) symmetric interaction, we also have a long-range one-pion-exchange (OPE) potential
	at leading order $\chi$EFT interaction. We define our one-pion-exchange potential following a recently developed regularization
	method~\cite{Reinert:2017usi},
	\begin{align}
	V_{\rm OPE}^{\Lambda_{\pi}}  = - &   \frac{g_A^2}{8f^2_{\pi}}\ \, \sum_{{\bf n',n},S',S,I}
	:\rho_{S',I\rm }^{(0)}(\vec{n}')f_{S',S}(\vec{n}'-\vec{n}) 
	\rho_{S,I}^{(0)}(\vec{n}):  
	\,,
	\label{eq:OPEP-full}
	\end{align}
	\begin{align}
	V_{\rm C_{\pi}}^{\Lambda_{\pi}}  = -C_{\pi} \, \frac{g_A^2}{8f^2_{\pi}} 
	\sum_{{\bf n',n},S,I}
	:\rho_{S,I}^{(0)}(\vec{n}')
	f^{\pi}(\vec{n}'-\vec{n})
	\rho_{S,I}^{(0)}(\vec{n}):\,
	.
	\label{eq:OPEP-counter}
	\end{align}
	Here $f^{\pi}$ is a local regulator in momentum space defined as
	\begin{align}
	f^{\pi}(\vec{n}'-\vec{n})
	= &
	\frac{1}{L^3}
	\sum_{\vec{q}}
	e^{-i\vec{q}\cdot(\vec{n}'-\vec{n})-(\vec{q}^2+M^2_{\pi})/\Lambda_{\pi}^2}\,,
	\end{align}
	$f_{S',S}$ is the locally-regulated pion correlation function,
	\begin{align}
	f_{S',S}(\vec{n}'-\vec{n}) 
	= &\frac{1}{L^3}\sum_{\vec{q}}
	\frac{q_{S'}q_{S} \, e^{-i\vec{q}\cdot(\vec{n}'-\vec{n})-(\vec{q}^2+M^2_{\pi})/\Lambda_{\pi}^2}}{\vec{q}^2 + M_{\pi}^2} \,,
	\end{align}
	and
	\begin{align}
	C_{\pi} = & -
	\frac{\Lambda_{\pi} (\Lambda_{\pi}^2-2M_{\pi}^{2}) + 2\sqrt{\pi} M_{\pi}^3\exp(M_{\pi}^2/\Lambda_{\pi}^{2}){\rm erfc}(M_{\pi}^2/\Lambda_{\pi}^{2})}
	{3 \Lambda_{\pi}^3}\,,
	\end{align}
	with $g_{A}=1.287$ the axial-vector coupling constant, $f_{\pi}=92.2$~MeV the pion decay constant and $M_{\pi}=134.98$~MeV the pion mass. The term given in Eq.~(\ref{eq:OPEP-counter}) is a counterterm introduced to remove the short-distance admixture in the one-pion-exchange potential~\cite{Reinert:2017usi}. In our simple Hamiltonian, we set  $\Lambda_{\pi}=180$~MeV and $C_{\pi} = 0$, and we compute the difference $V_{\rm OPE}^{\Lambda_{\pi}=300}-V_{\rm OPE}^{\Lambda_{\pi}=180}$ and the OPEP counterterm $V_{\rm C_{\pi}}^{\Lambda_{\pi}}$ perturbatively. Here we use the notation
	\begin{align}
	\rho^{(d)}(\vec{n}) = \sum_{i,j=0,1} 
	a^{\dagger}_{i,j}(\vec{n}) \, a^{\,}_{i,j}(\vec{n})
	+
	s_{\rm L}
	\sum_{|\vec{n}-\vec{n}^{\prime}|^2 = 1}^d 
	\,
	\sum_{i,j=0,1} 
	a^{\dagger}_{i,j}(\vec{n}^{\prime}) \, a^{\,}_{i,j}(\vec{n}^{\prime})
	.
	\label{eqn:appx--001}
	\end{align}
	and 
	\begin{align}
	\rho^{(d)}_{S,I}(\vec{n}) = & \sum_{i,j,i^{\prime},j^{\prime}=0,1} 
	a^{\dagger}_{i,j}(\vec{n}) \, [\sigma_{S}]_{ii^{\prime}} \, [\sigma_{I}]_{jj^{\prime}} \, a^{\,}_{i^{\prime},j^{\prime}}(\vec{n})
	\nonumber \\
	& +
	s_{\rm L}
	\sum_{|\vec{n}-\vec{n}^{\prime}|^2 = 1}^d 
	\,
	\sum_{i,j,i^{\prime},j^{\prime}=0,1} 
	a^{\dagger}_{i,j}(\vec{n}^{\prime}) \, [\sigma_{S}]_{ii^{\prime}} \, [\sigma_{I}]_{jj^{\prime}} \, a^{\,}_{i^{\prime},j^{\prime}}(\vec{n}^{\prime})
	\label{eqn:appx--005}
	\end{align}
	for the density operators. 
	
	\subsubsection{Hamiltonian at {\rm N3LO} (next-to-next-to-next-to-leading order)}

	We now give the details of our realistic Hamiltonian $H$. Let us define the functions
	\begin{align}
	f_{S}^{\pi}(\vec{n}^{\prime}-\vec{n})
	=
	\frac{1}{L^3} 
	\sum_{\vec{q}}
	e^{-i\vec{q}\cdot(\vec{n}'-\vec{n})-(\vec{q}^2+M^2_{\pi})/\Lambda_{\pi}^2}
	\,
	q_{S}
	\label{eqn:appx--0013}
	\end{align}
	and
	\begin{align}
	f_{S}^{\pi\pi}(\vec{n}^{\prime}-\vec{n})
	=
	\frac{1}{L^{3}} \sum_{\vec{q}}
	\frac{ e^{-i\vec{q}\cdot(\vec{n}'-\vec{n})-(\vec{q}^2+M^2_{\pi})/\Lambda_{\pi}^2} }
	{\vec{q}^{2}+ M_{\pi}^{2}}
	\, q_{S}
	\,.
	\label{eqn:appx--0013prime}
	\end{align}
	We define the Hamiltonian $H$ using $\chi$EFT interactions at N3LO,
	\begin{equation}
	H = K 
	+ V_{\rm OPE}^{\Lambda_{\pi}} 
	+ V_{\rm C_{\pi}}^{\Lambda_{\pi}} 
	+ V_{\rm Coulomb}
	+ V_{\rm 3N}^{\rm Q^3}
	+ V_{\rm 2N}^{\rm Q^4}
	+ W_{\rm 2N}^{\rm Q^4}
	+ V_{\rm 2N,WFM}^{\rm Q^4}
	+ W_{\rm 2N,WFM}^{\rm Q^4},
	\label{eq:H-N3LO}
	\end{equation}
	where $K$ is the kinetic energy term. $V_{\rm OPE}^{\Lambda_{\pi}}$ and $V_{\rm C_{\pi}}^{\Lambda_{\pi}}$ are defined in Eqs.~(\ref{eq:OPEP-full}) and (\ref{eq:OPEP-counter}) with ${\Lambda_{\pi}=300}$~MeV. $V_{\rm Coulomb}$ is the Coulomb interaction, $V_{\rm 3N}^{\rm Q^3}$ is the 3N potential, $V_{\rm 2N}^{\rm Q^4}$ is the 2N short-range interaction at N3LO of $\chi$EFT,  $W_{\rm 2N}^{\rm Q^4}$ is the 2N Galilean invariance restoration (GIR) interaction at N3LO of $\chi$EFT, $V_{\rm 2N,WFM}^{\rm Q^4}$ is the wave function matching interaction defined as $H^{\prime}-H$, and $W_{\rm 2N,WFM}^{\rm Q^4}$ is the GIR correction of the wave function matching interaction.

	For the details of the Coulomb interaction and the 2N short-range interactions we refer the reader to Ref.~\cite{Li:2018ymw}. The three-nucleon interactions at ${\rm Q^3}$ consists of a contact potential, one-pion exchange potential, and two-pion exchange potential~\cite{Friar:1998zt,Epelbaum:2002vt,Epelbaum:2009zsa}, and in this work we defined two additional SU(4) symmetric potentials denoted by $V_{c_{E}}^{(l)}$ and $V_{c_{E}}^{(t)}$. Therefore, the three-nucleon interactions at ${\rm Q^3}$ has the form
	\begin{align}
	V_{\rm 3N}^{\rm Q^3}
	= V_{c_{E}}^{(l)}
	+ V_{c_{E}}^{(t)}
	+ V_{c_{E}}^{(d)}
	+ V_{c_{D}}^{(d)}
	+ V_{\rm 3N}^{\rm (TPE)}
	\,,
	\label{eqn:V_NNLO^3N--001}
	\end{align}
	where
	\begin{align}
	V_{c_{E}}^{(d)}
	= 
	\frac{1}{6}
	\frac{c_{E}^{(d)}}{2 f_{\pi}^{4} \Lambda_{\chi}} 
	\, : \,
	\sum_{\vec{n}} 
	\left[
	\rho^{(d)}(\vec{n})
	\right]^3
	\, : \,
	\,,
	\label{eqn:V_contact^3N--001}
	\end{align}
	\begin{align}
	V_{c_{D}}^{(d)}
	= -\frac{c_{D}^{(d)} \, g_{A}}{4f_{\pi}^{4} \Lambda_{\chi}}\,  \sum_{\vec{n},S,I}
	\sum_{\vec{n}^{\prime},S^{\prime}}
	\, : \,
	\rho^{(0)}_{S^{\prime},I}(\vec{n}^{\prime})
	f_{S^{\prime},S}(\vec{n}^{\prime}-\vec{n})
	\rho^{(d)}_{S,I}(\vec{n})
	\rho^{(d)}(\vec{n})
	\, : \,
	\,,
	\label{eqn:V_OPE^3N--001}
	\end{align}
	
	\begin{align}
	V_{c_{E}}^{(l)}= c_{E}^{(l)}
	\, \sum_{\vec{n},\vec{n}^{\prime},\vec{n}^{\prime\prime}}
	\rho^{(0)}(\vec{n})\,
	\rho^{(0)}(\vec{n}^{\prime}) \, 
	\rho^{(0)}(\vec{n}^{\prime\prime}) \delta_{|\vec{n}-\vec{n}^{\prime}|,1} \, \,  \delta_{|\vec{n}-\vec{n}^{\prime\prime}|,1} \, \,  \delta_{|\vec{n}^{\prime}-\vec{n}^{\prime\prime}|,2},               
	\end{align}
	\begin{align}
	V_{c_{E}}^{(t)} = c_{E}^{(t)}                \,
	\sum_{\vec{n},\vec{n}^{\prime},\vec{n}^{\prime\prime}}
	\rho^{(0)}(\vec{n})\,
	\rho^{(0)}(\vec{n}^{\prime}) \, 
	\rho^{(0)}(\vec{n}^{\prime\prime}) \delta_{|\vec{n}-\vec{n}^{\prime}|,\sqrt{2}} \, \,  \delta_{|\vec{n}-\vec{n}^{\prime\prime}|,\sqrt{2}} \, \,  \delta_{|\vec{n}^{\prime}-\vec{n}^{\prime\prime}|,\sqrt{2}}. 
	\end{align}
	The $V_{\rm 3N}^{\rm (TPE)}$ potential can be separated into the following three parts,
	\begin{align}
	V_{\rm 3N}^{\rm (TPE1)}
	= & \frac{c_{3}}{f_{\pi}^{2}}\,
	\frac{g_{A}^{2}}{4 f_{\pi}^{2}}
	\, 
	\sum_{S,S^{\prime},S^{\prime\prime},I}
	\sum_{\vec{n},\vec{n}^{\prime},\vec{n}^{\prime\prime}}
	\nonumber \\
	&
	\times	
	\, : \, 
	\rho^{(0)}_{S^{\prime},I}(\vec{n}^{\prime}) \,
	f_{S^{\prime},S}(\vec{n}^{\prime}-\vec{n})
	f_{S^{\prime\prime},S}(\vec{n}^{\prime\prime}-\vec{n})
	\rho^{(0)}_{S^{\prime\prime},I}(\vec{n}^{\prime\prime}) \,
	\rho^{(0)}(\vec{n})
	\, : \,
	\label{eqn:V_TPE1^3N--001}
	\end{align}
	
	\begin{align}
	V_{\rm 3N}^{\rm (TPE2)}
	= &
	-\frac{2c_{1}}{f_{\pi}^{2}}\,
	\frac{g_{A}^{2} \, M_{\pi}^{2}}{4 f_{\pi}^{2}}
	\, 
	\sum_{S,S^{\prime},I}
	\sum_{\vec{n},\vec{n}^{\prime},\vec{n}^{\prime\prime}}
	\nonumber \\
	&
	\times	
	\, : \, 
	\rho^{(0)}_{S^{\prime},I}(\vec{n}^{\prime}) \,
	f_{S^{\prime}}^{\pi\pi}(\vec{n}^{\prime}-\vec{n})
	f_{S}^{\pi\pi}(\vec{n}^{\prime\prime}-\vec{n})
	\rho^{(0)}_{S,I}(\vec{n}^{\prime\prime}) \,
	\rho^{(0)}(\vec{n})
	\, : \, 
	\,,
	\label{eqn:V_TPE2^3N--001}
	\end{align}
	
	\begin{align}
	V_{\rm 3N}^{\rm (TPE3)}
	=   \frac{c_{4}}{2f_{\pi}^{2}}
	&
	\left( \frac{g_{A}}{2 f_{\pi}}\right)^{2}
	\sum_{S_{1},S_{2},S_{3}}
	\sum_{I_{1},I_{2},I_{3}}
	\sum_{S^{\prime},S^{\prime\prime}}
	\sum_{\vec{n},\vec{n}^{\prime},\vec{n}^{\prime\prime}}
	\varepsilon_{S_1,S_2,S_3}
	\varepsilon_{I_1,I_2,I_3}
	\nonumber \\
	&
	\times	
	\, : \, 
	\rho^{(0)}_{S^{\prime},I_{1}}(\vec{n}^{\prime}) \,
	f_{S^{\prime},S_{1}}(\vec{n}^{\prime}-\vec{n})
	f_{S^{\prime\prime},S_{2}}(\vec{n}^{\prime\prime}-\vec{n})
	\rho^{(0)}_{S^{\prime\prime},I_{2}}(\vec{n}^{\prime\prime}) \,
	\rho^{(0)}_{S_{3},I_{3}}(\vec{n})
	\, : \, \,.
	\label{eqn:V_TPE3^3N--001}
	\end{align}

\begin{figure}[htb!]
		\centering\includegraphics[width=1.0\textwidth]{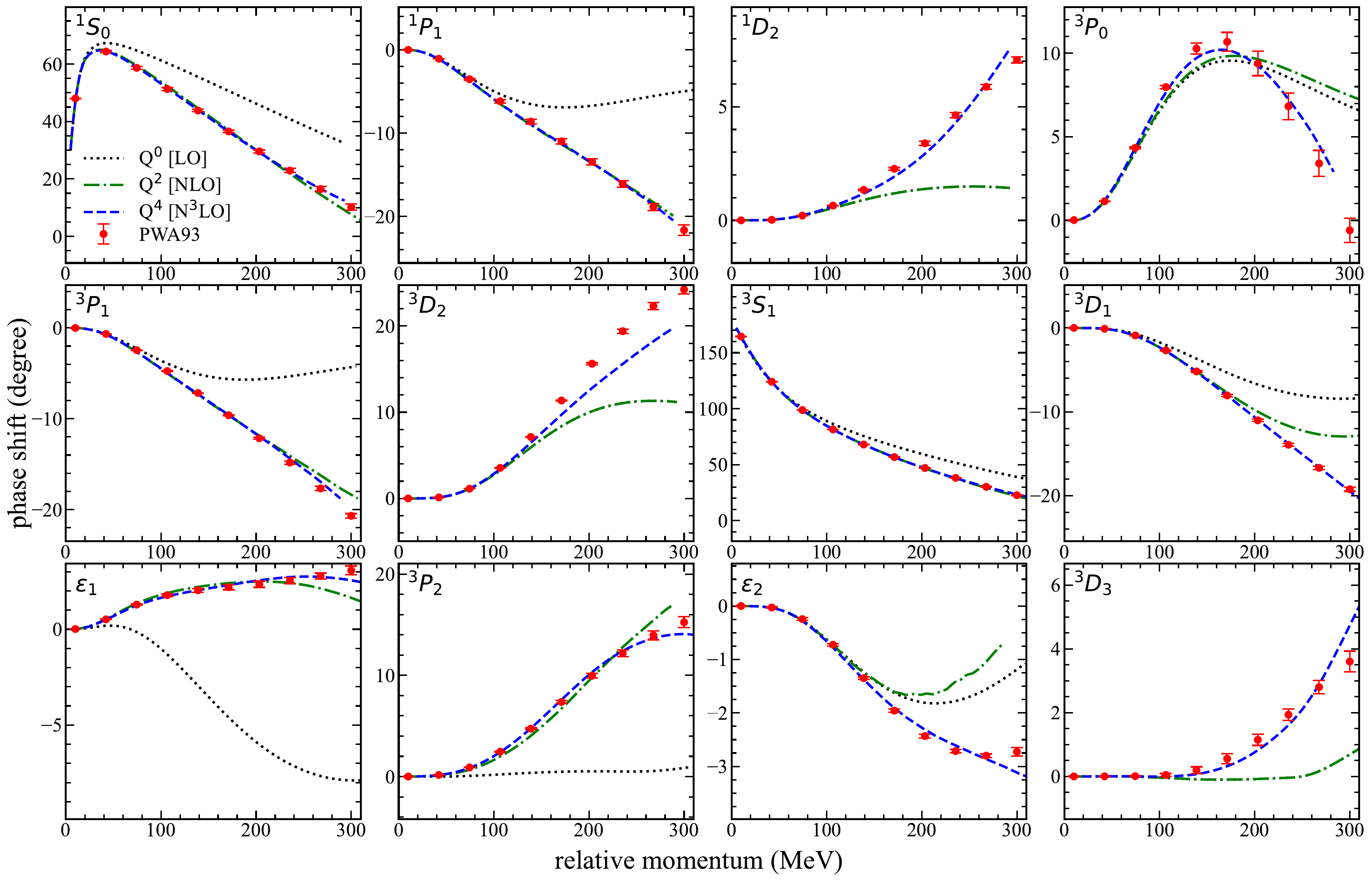}
		\caption{Plots of the neutron-proton scattering phase shifts as functions of relative momenta with lattice spacing $a=1.32$~fm. For comparison we also plot the phase shifts extracted from the Nijmegen partial wave analysis~\cite{Stoks:1993tb}.}
		\label{fig:NN-phaseshifts}
	\end{figure}

	We perform our calculations using lattice spacing $a = 1.32$~fm, and we determine the low-energy constants (LECs) of the 2N short-range interaction up to N3LO of $\chi$EFT by reproducing the neutron-proton scattering phase shifts and mixing angles of the Nijmegen partial wave analysis (PWA)~\cite{Stoks:1993tb}. The lattice spacing of $a=1.32$~fm corresponds to the momentum space cut-off of $470$~MeV, which corresponds to the resolution scale, at which the hidden spin-isospin symmetry of the
	NN interactions is best fulfilled~\cite{Lee:2020esp}. In Fig.~\ref{fig:NN-phaseshifts} we plot the calculated neutron-proton scattering phase shifts up to N3LO of $\chi$EFT as functions of relative momenta with comparison to the Nijmegen PWA. Only the statistical errors and not systematic errors are included in the Nijmegen PWA. In the next subsection we discuss the approach used to estimate the uncertainties in our calculations and the determination of the LECs of the three-nucleon interactions.
	
	\subsection{Uncertainty Analysis}
	
	We discuss the analysis and quantify the errors which include relevant sources of uncertainty in the calculations presented in this paper. To perform such a complete analysis to estimate the uncertainties for the calculations using chiral interactions, a global parameter search for the LECs of the chiral interactions is required. This task could be extremely difficult as one has to perform a search over a high-dimensional parameter space. Nevertheless, by performing some prior analyses we can reduce both the dimension and the volume of the parameter space to be searched. For instance, the LECs of the two-pion exchange three-body potentials are already fixed from pion--nucleon scattering data, $c_{1}=-1.10(3)$, $c_{3}=-5.54(6)$ and $c_{4}=4.17(4)$ all in GeV$^{-1}$~\cite{Hoferichter:2015tha}, and the LECs of two-nucleon potentials can be constrained using the empirical partial wave phase shifts and mixing angles, which help to shrink the parameter space. However, the main difficulty is the determination of the unknown LECs of the three--nucleon forces. As we discuss in the following, we treat these LECs as unknown regression coefficients of an emulator employed with history matching, which has been shown to be an effective approach for parameter searches~\cite{10.1214/10-BA524,10.1214/12-STS412,Hu:2021trw}.

\begin{figure}[htb]
		\centering\includegraphics[width=1.0\textwidth]{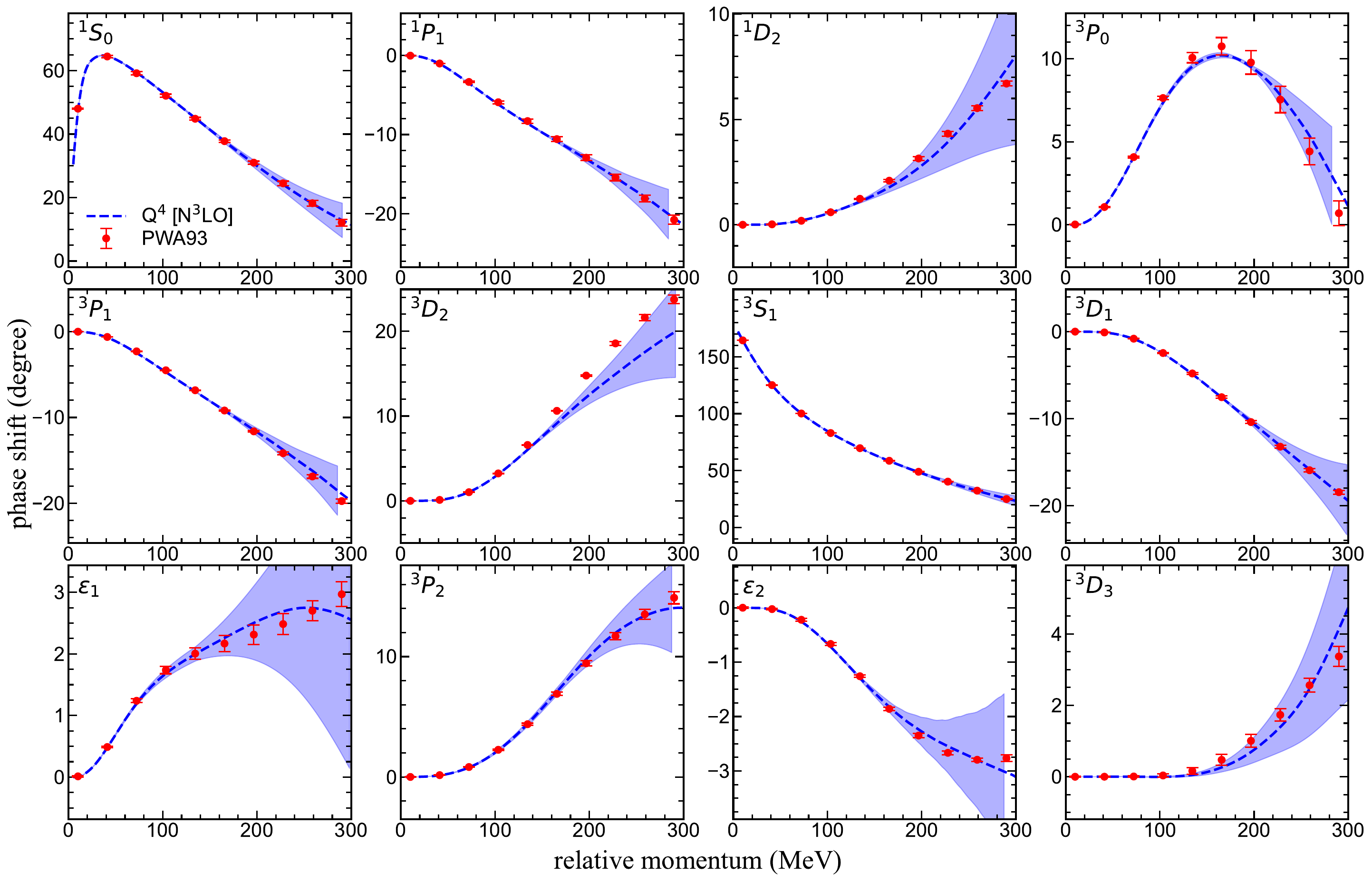}
		\caption{The plot of the theoretical error bands for the neutron-proton scattering phase shifts and mixing angles versus the relative momenta.}
		\label{fig:np-phase-shifts-w-errors}
	\end{figure}

	In our calculations we determine the LECs of the 2N chiral interactions by fitting the calculated neutron-proton scattering phase shifts and mixing angles on the lattice to the Nijmegen PWA~\cite{Stoks:1993tb}. To estimate the systematic theoretical uncertainties, we use the method introduced in Refs.~\cite{Epelbaum:2014efa,Epelbaum:2014sza} and calculate the theoretical errors for the neutron-proton scattering phase shifts and mixing angles as a function of the relative momenta, and the results are shown in Fig.~\ref{fig:np-phase-shifts-w-errors}.
	One can find detailed discussions about the convergence of the chiral effective field theory expansion on the lattice and theoretical errors in Ref.~\cite{Li:2018ymw}. As seen in Fig.~\ref{fig:np-phase-shifts-w-errors} the theoretical error bands at N3LO give a considerable-sized parameter space of LECs. Therefore, we use history matching together with an emulator to filter the 2N LECs as well as to identify the 3N LECs which provide an acceptable matching between {\it ab initio} calculations and the experimental data for some selected nuclei.

\begin{figure}[htb]
		\centering\includegraphics[width=1.0\textwidth]{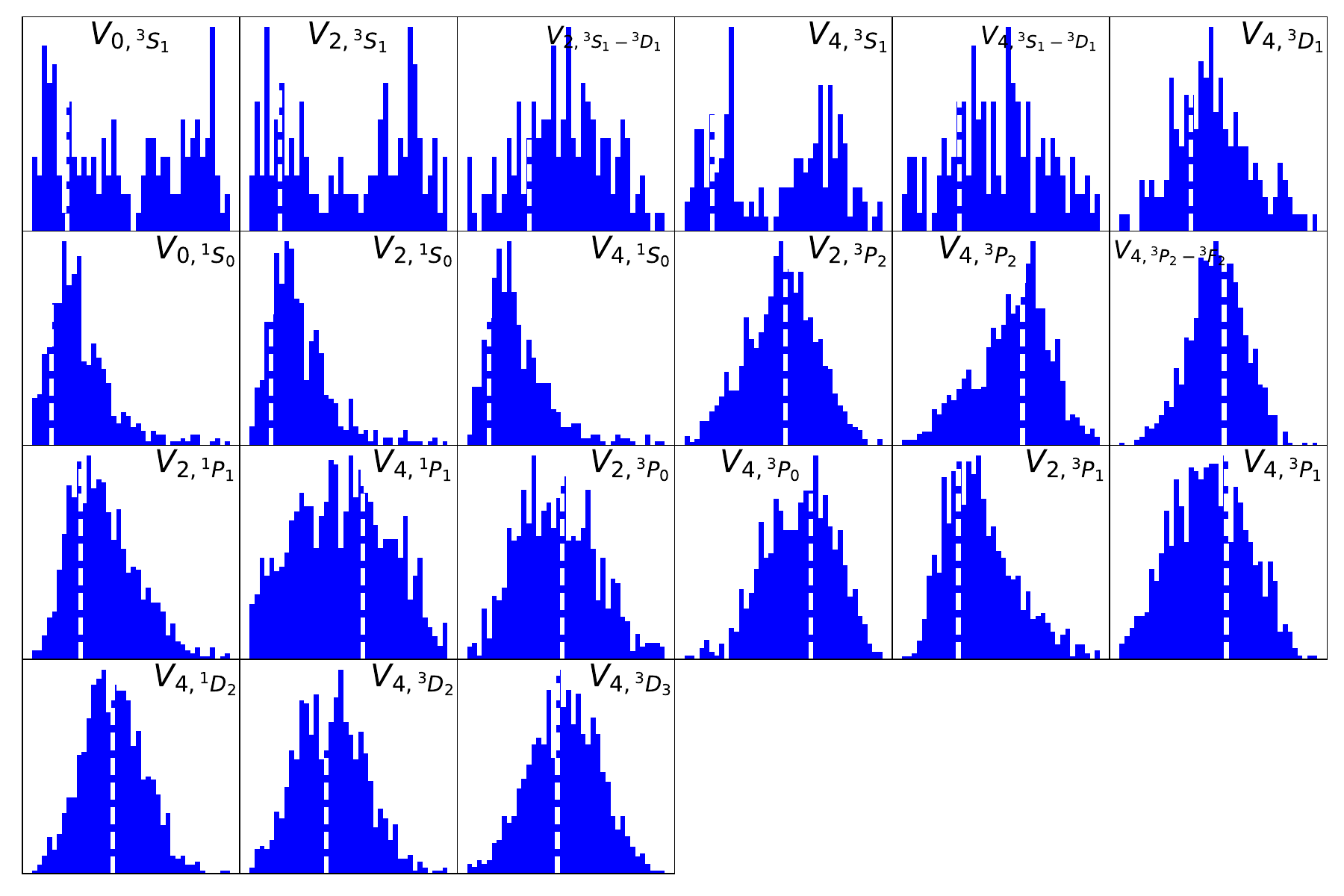}
		\caption{The plot of the posterior distribution of the 2N LECs at N3LO sampled using MCMC method.}
		\label{fig:2N-LECs-Dist.}
	\end{figure}

	The first step in our analysis is to use the theoretical errors shown in Fig.~\ref{fig:np-phase-shifts-w-errors} and get posterior distributions of the 2N LECs sampling with Markov Chain Monte Carlo (MCMC) and the results are shown in Fig.~\ref{fig:2N-LECs-Dist.}.
	For each interaction, we approximately collect $10^3$ samples. Now we can use 2N LECs from these distributions as inputs into our {\it ab-initio} model, $f(x)$, for which we define the emulator equation given by 
	\begin{align}
	f_{o}(\mathbf{x}) = 
	\sum_{i} x_{i} \left(\frac{\partial E}{\partial  x}\right)_{io}
	+ \sum_{k} \beta_{k} \left(\frac{\partial E}{\partial \beta}\right)_{ko}\,.
	\label{eqn:emaulator-eqn}
	\end{align}
	The first term on the right hand side is the model output of only 2N interactions using 2N inputs, which is denoted by the vector $\mathbf{x}$, from the distributions shown in Fig.~\ref{fig:2N-LECs-Dist.}. The second term on the right hand side is the output of only 3N interactions and it is a calibration term where $\beta_{k}$ are unknown regression coefficients which will correspond to the 3N LECs at the end of the analysis. For this term we consider five different modified forms of the SU(4) symmetric three-nucleon forces and three different modified forms of the one-pion-exchange three-nucleon forces at short distances, $V_{c_{E}}^{(0)}$, $V_{c_{E}}^{(1)}$, $V_{c_{E}}^{(2)}$, $V_{c_{E}}^{(l)}$, $V_{c_{E}}^{(t)}$, $V_{c_{D}}^{(0)}$, $V_{c_{D}}^{(1)}$ and $V_{c_{D}}^{(2)}$. In the emulator Eq.~(\ref{eqn:emaulator-eqn}), the derivatives of the energies with respect to $x$ and $\beta$ stand for the calculated operators on the lattice using first-order perturbation theory. Also, in the emulator equation all inputs are active variables, therefore, we do not make any such distinction. The emulator works for any number of outputs, $f_{o}(x)$ with $o = 1,2,\ldots,q$. When we emulate our model's behavior for $x_i$ inputs, with $i = 1,\cdots,d$, we use the experimental data corresponding to the outputs and the calibration coefficients $\beta_{k}$ so that there is acceptable agreement between our model and the experimental data. Therefore, the calibration coefficients $\beta_{k}$ are obtained from a nonlinear least-squares fitting by minimizing the expression for each input set in the vector $\mathbf{x}$,
	\begin{align}
	r_{n}(\mathbf{x}) = \left[\min \left(f_{n}(\mathbf{x}) -z_{n}\right)^2 \right]^{1/2}\,,
	\label{eqn:non-lin-least-sq}
	\end{align}
	where $z_{n}$ is the experimental data. When we evaluate Eq.~(\ref{eqn:non-lin-least-sq}), we impose a constraint on the least square problem so that only natural sized parameters are allowed for the coefficient $\beta_{k}$.
	
	In the second step, we use Eq.~(\ref{eqn:non-lin-least-sq}) to link our model to reality by evaluating inputs from the distributions shown in Fig.~\ref{fig:2N-LECs-Dist.} and determining the 3N LECs. To achieve this, we use history matching to perform an efficient parameter search. History matching is an iterative process that identifies and eliminates implausible parts of the input space by measuring implausibility of inputs, shrinks the input space in every iteration, and repeats the non-implausible input search in the smaller input space. Here we use $\mathcal{Q}_{w}$ to denote the volume of non-implausible input space. The implausibility measure of a given $d$ inputs $\mathbf{x} = (x_{1},x_{2},\cdots,x_{d})$ for an $n^{\rm th}$ output is written as,
	\begin{align}
	I_{n}(\mathbf{x}) 
	= \frac{r_{n}(\mathbf{x})}
	{\sqrt{{\rm Var}\left[r_{n}(\mathbf{x})\right]+{\rm Var}\left[\epsilon_{n}\right]}}\,,
	\label{eqn:implausibility-measure}
	\end{align}
	where the numerator is the distance between the experimental data $z_{n}$ and the calculated output $f_{n}(\mathbf{x})$, and the denominator is the square root of the sum of the variance of the distance $r_{n}(\mathbf{x})$ and the variance of the errors of prediction $\epsilon_{n}$. Any particular value of $\mathbf{x}$, say $x_{i}$, that give a large value of $I_{n}(\mathbf{x})$ can be considered as implausible which means that for this input it is unlikely to get an acceptable match between the calculated outputs and the experimental data. Therefore, we can use a maximum implausibility measure $I_{M}(\mathbf{x})$ and discard the input from the input space provided that $I_{M}(\mathbf{x}) > c$ where $c$ is a cutoff for non-implausibility. In our analysis we define the maximum implausibility measure by maximizing over all outputs,
	\begin{align}
	I_{M}(\mathbf{x}) 
	= \max_{o \,  \in \,  \mathcal{Z}_{w}}I_{o}(\mathbf{x})\,,
	\label{eqn:max-implausibility-measure}
	\end{align}
	where $\mathcal{Z}_{w}$ is the set of outputs that we consider in the $w^{\rm th}$ iteration, known as wave, and we use $c = 3$ by appealing to Pukelsheim's 3--sigma rule~\cite{pukelsheim1994three} which indicates that 95\% of the all population lies within $\pm 3\sigma$. Even though the maximum implausibility measure is defined by Eq.~(\ref{eqn:max-implausibility-measure}), in our analysis we look at the first three maximum implausibilities in every iteration, and it provides us a tool to select three-nucleon forces which work as a calibrator in Eq.~(\ref{eqn:implausibility-measure}). This is a crucial step both as it gives the information about three-nucleon forces which are inevitable to obtain an acceptable agreement between the calculated outputs and the experimental data with some reasonable credibility and as it guides us to not discard any useful information in the input space in early iterations.
	
	For the lattice Monte Carlo simulations, the general strategy is to use the expansion $H' = H^S + (H' - H^S)$ and apply corrections up to first order in perturbation theory.  In order to accelerate the convergence of perturbation theory further, we find it helpful to consider the more general partition $H' = H'^S + (H' - H'^S)$.  The modified simple Hamiltonian $H'^S$ has the same form as $H^S$ in Eq.~(\ref{eq:Hsimple2}), but we allow for different coupling strengths $c_{\rm SU(4)}$ and $c_{I}$.  We then minimize the energy and use the variational principle to optimize the parameters.  
	
	We refer the readers to Ref.~\cite{vernon2018bayesian} for a general description of and a recipe for history matching and to Ref.~\cite{Hu:2021trw} for a nuclear physics application. In the following we discuss history matching method and outline the detailed procedure of its application to our model. 
	
	\begin{description}
		\item[WAVE 1\label{itm:wave-1}] In the first wave, the implausibility measure is run over outputs $\mathcal{Z}_{1}=\{{}^3{\rm H},{}^4{\rm He}\}$ and inputs $\mathbf{x} =\mathbf{x}_{{}^1S_{0}} \otimes \mathbf{x}_{{}^3S_{1}-{}^3D_1}$. Our choice for the collection of inputs in the volume $\mathcal{Q}_{0}$ is based on the fact that only interactions with zero angular momentum are giving significant contributions to the nuclei used as informative observables for this first wave. Therefore, we define the current sets of non-implausible inputs $\mathbf{x}$ from the distributions shown in Fig.~\ref{fig:2N-LECs-Dist.}, and we calculate the model outputs from Eq.~(\ref{eqn:emaulator-eqn}) by performing a calibration using least square regression over the second term on the right hand side with the simplest 3N interactions $V_{c_{E}}^{(0)}$ and $V_{c_{D}}^{(0)}$. Then, the implausibility measures $I_{M}(\mathbf{x})$ are calculated using the first maximum implausibility over outputs, and implausibility cutoffs are imposed to define the new non-implausible volume denoted by $\mathcal{Q}_{1}$. The results of the fitted (predicted) observables are shown by black open (filled) down-triangle points in Fig.~\ref{fig:HM-waves-Epred} where we plots the deviations between the model outputs and experimental data given in percentage. From Fig.~\ref{fig:HM-waves-Epred} it can be seen clearly that when we run the emulator for outputs $\mathcal{Z}_{1}$, deviations for all predicted outputs are negative, which is important information for choosing outputs to be included in the emulator in the next iteration. 
		\begin{figure}[htb]
			\centering\includegraphics[width=1.0\textwidth]{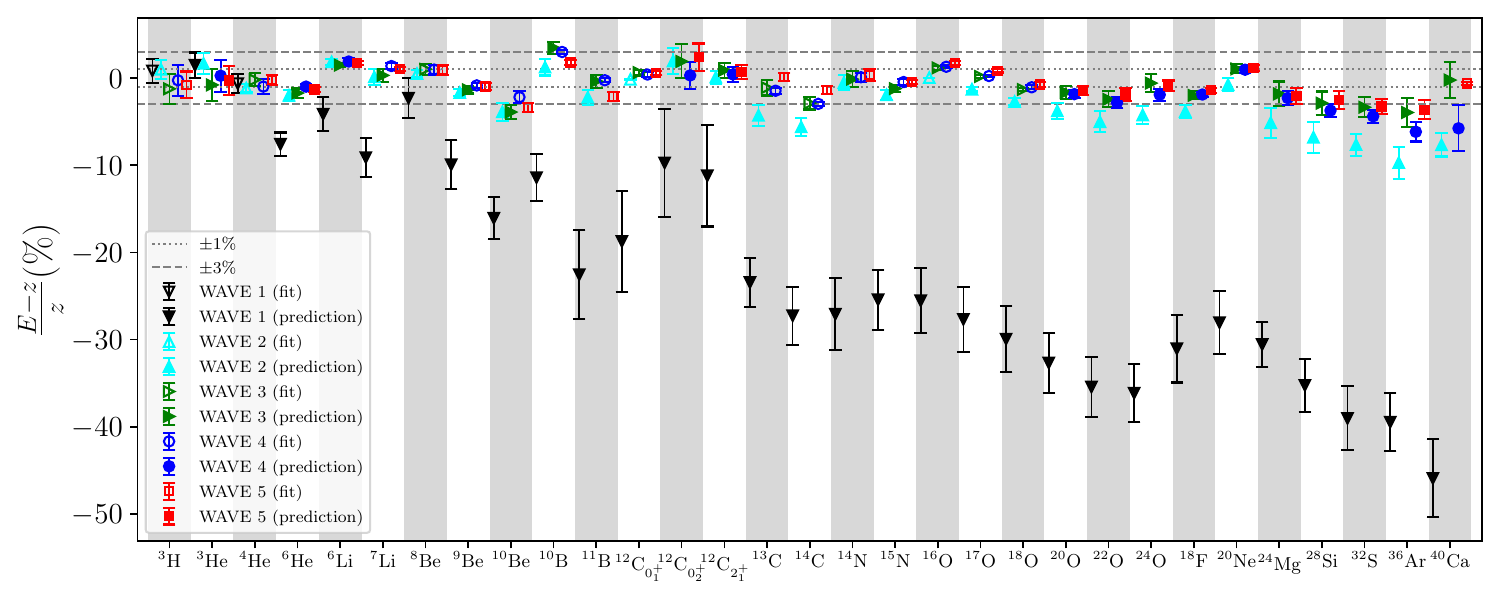}
			\caption{The plots of deviations between the model outputs and experimental data per nucleon at history matching iterations from \ref{itm:wave-1} to \ref{itm:wave-5}.}
			\label{fig:HM-waves-Epred}
		\end{figure}
		\item[WAVE 2\label{itm:wave-2}]  At this wave, we run the implausibility measure over outputs $\mathcal{Z}_{2} =$ $\mathcal{Z}_{1}$ $\, \bigcup \, $ $\{{}^{8}{\rm Be}$, ${}^{12}{\rm C}$, ${}^{16}{\rm O}$$\}$ by considering the fact that there exist strong evidences that the structures of these additional nuclei are in the form of distinct geometrical configuration of alpha clusters, then we evaluate the inputs for all 2N interactions. The current non-implausible parameter space for the ${}^1S_{0}$ and ${}^3S_{1}-{}^3D_1$ channels are set by \ref{itm:wave-1}, while for the channels that have not been evaluated yet the current sets of non-implausible parameter space is defined from the distributions shown in Fig.~\ref{fig:2N-LECs-Dist.}. At this wave we perform the least square regression given in Eq.~(\ref{eqn:non-lin-least-sq}) by using two additional 3N interactions. To make selection for these additional 3N interactions we run the emulator for all possible sets of 3N interactions and compute the first three non-zero maximum implausibility measures. Sets of 3N interaction used as calibrators and the corresponding results for the ratio of the non-implausible space volumes, $\mathcal{Q}_{2}/\mathcal{Q}_{1}$, are shown in Table~\ref{tab:non-imp-wave2}.	
		\begin{table}[ht]
			\centering
			\caption{Results for the first three non-zero maximum implausibility measures at \ref{itm:wave-2}. We run the implausibility measure over outputs ${}^3{\rm H}$, ${}^4{\rm He}$,${}^{8}{\rm Be}$,${}^{12}{\rm C}$, and ${}^{16}{\rm O}$.}
			\begin{tabular}{ |L{4cm} || C{2cm} | C{2cm} | C{2cm} | }
				\hline\hline
				3N interactions &  \multicolumn{3}{|c|}{ratio of non-implausible input volumes $\mathcal{Q}_{2}/\mathcal{Q}_{1}$ (\%)} \\ 
				in addition to $V_{c_{E}}^{(0)}$, $V_{c_{D}}^{(0)}$ & $I^{(1)}_{M}(\mathbf{x})$ &  $I^{(2)}_{M}(\mathbf{x})$ &  $I^{(3)}_{M}(\mathbf{x})$ \\
				\hline\hline
				$V_{c_{E}}^{(l)}$,  $V_{c_{E}}^{(t)}$
				&~~42    &95&  99\\\hline
				$V_{c_{E}}^{(2)}$,  $V_{c_{E}}^{(t)}$
				&~~33    &94&   99\\\hline
				$V_{c_{E}}^{(2)}$,  $V_{c_{E}}^{(l)}$
				&~~30       &95&   99\\\hline
				$V_{c_{E}}^{(1)}$,  $V_{c_{E}}^{(t)}$
				&$<1$     &37&   97\\\hline
				$V_{c_{E}}^{(1)}$,  $V_{c_{E}}^{(l)}$
				&$<1$     &26&   86\\\hline
				$V_{c_{D}}^{(2)}$,  $V_{c_{E}}^{(t)}$
				&$<1$     &20&   97\\\hline
				$V_{c_{D}}^{(1)}$,  $V_{c_{E}}^{(l)}$
				&$<1$     &14&   80\\\hline
				$V_{c_{E}}^{(1)}$,  $V_{c_{E}}^{(2)}$  
				&$<1$     &11&   88\\\hline
				$V_{c_{D}}^{(2)}$,  $V_{c_{E}}^{(l)}$
				&$<1$     &10&   82\\\hline
				$V_{c_{D}}^{(1)}$,  $V_{c_{D}}^{(2)}$
				&$<1$     &10&   70\\\hline
				$V_{c_{E}}^{(2)}$,  $V_{c_{D}}^{(1)}$
				&$<1$     &~7&   92\\\hline
				$V_{c_{E}}^{(1)}$,  $V_{c_{D}}^{(1)}$
				&$<1$     &~2&   39\\\hline
				$V_{c_{E}}^{(1)}$,  $V_{c_{D}}^{(2)}$
				&$<1$     &~2&   32\\\hline
				$V_{c_{D}}^{(1)}$,  $V_{c_{E}}^{(t)}$
				&$<1$     &~1&   97\\\hline
				$V_{c_{E}}^{(2)}$,  $V_{c_{D}}^{(2)}$
				&$<1$     &~1&   96\\\hline
			\end{tabular}
			\label{tab:non-imp-wave2}
		\end{table}

		As seen from the results, in the most cases one of the calculated outputs has a large deviation with small variance, and as a result the first maximum implausibility measures are less than 1\%, and in the cases where two additional 3N interactions are $V_{c_{E}}^{(l)}$ and $V_{c_{E}}^{(t)}$ we obtain the largest non-implausible space volume. Therefore, from Table~\ref{tab:non-imp-wave2} we find the most promising set of 3N interaction as  $V_{c_{E}}^{(0)}$, $V_{c_{D}}^{(0)}$, $V_{c_{E}}^{(l)}$ and $V_{c_{E}}^{(t)}$ shown in the first row of Table~\ref{tab:non-imp-wave2}. Therefore, we define the new non-implausible space volume $\mathcal{Q}_{2}$ at this iteration. 
		
		\item[WAVE 3\label{itm:wave-3}]  The results of the fitted (predicted) observables at \ref{itm:wave-2} are shown by cyan open (filled) up-triangle points in Fig.~\ref{fig:HM-waves-Epred}. We find that using four 3N interactions in \ref{itm:wave-3} already gives a decent description for nuclei with $N=Z=8$, while deviations for neutron-rich nuclei with $Z\le8$ are large. Therefore, at this wave we run the the implausibility measure over outputs $\mathcal{Z}_{3} =$ $\mathcal{Z}_{2}$ $\, \bigcup \, $ $\{{}^{13}{\rm C}$, ${}^{14}{\rm C}$, ${}^{17}{\rm O}$, ${}^{18}{\rm O}$$\}$ and we evaluate the inputs for all 2N interactions. The current non-implausible input volume is set by \ref{itm:wave-2}, $\mathcal{Q}_{2}$. Also, we run Eq.~(\ref{eqn:non-lin-least-sq}) by using two more 3N interactions in addition to $V_{c_{E}}^{(0)}$, $V_{c_{D}}^{(0)}$, $V_{c_{E}}^{(l)}$ and $V_{c_{E}}^{(t)}$. To select two additional 3N interactions out of four we run the emulator for all possible sets of 3N interactions and compute the first three non-zero maximum implausibility measures. Sets of 3N interaction used as calibrators and the corresponding results for the ratio of the non-implausible space volumes, $\mathcal{Q}_{3}/\mathcal{Q}_{2}$, are shown in Table~\ref{tab:non-imp-wave3}.	
		
		\begin{table}[ht]
			\centering
			\caption{Results for the first three non-zero maximum implausibility measures at \ref{itm:wave-3}. We run the implausibility measure over outputs ${}^3{\rm H}$, ${}^4{\rm He}$, ${}^{8}{\rm Be}$, ${}^{12}{\rm C}$, ${}^{13}{\rm C}$, ${}^{14}{\rm C}$, ${}^{16}{\rm O}$,  ${}^{17}{\rm O}$, and ${}^{18}{\rm O}$.}
			\begin{tabular}{ |L{4.7cm} || C{2cm} | C{2cm} | C{2cm} | }
				\hline\hline
				3N interactions  &  \multicolumn{3}{c|}{ratio of non-implausible input volumes $\mathcal{Q}_{3}/\mathcal{Q}_{2}$ (\%)} \\ 
				in addition to $V_{c_{E}}^{(0)}$, $V_{c_{D}}^{(0)}$, $V_{c_{E}}^{(l)}$, $V_{c_{E}}^{(t)}$& $I^{(1)}_{M}(\mathbf{x})$ &  $I^{(2)}_{M}(\mathbf{x})$ &  $I^{(3)}_{M}(\mathbf{x})$ \\
				\hline\hline
				$V_{c_{E}}^{(1)}$, $V_{c_{D}}^{(1)}$
				& ~~16  & ~~49   &90\\\hline
				$V_{c_{E}}^{(1)}$, $V_{c_{D}}^{(2)}$
				& ~~~~7  & ~~30   &92\\\hline
				$V_{c_{E}}^{(2)}$, $V_{c_{D}}^{(1)}$
				& ~~~~2  & ~~80   &93\\\hline
				$V_{c_{E}}^{(2)}$, $V_{c_{D}}^{(2)}$
				&  $<1$  &~~ 82   &98\\\hline
				$V_{c_{D}}^{(1)}$, $V_{c_{D}}^{(2)}$
				&  $<1$   & ~~53  &76\\
				\hline
				$V_{c_{E}}^{(1)}$, $V_{c_{E}}^{(2)}$
				&  $<1$  & $<1$   &~~6\\\hline\hline
			\end{tabular}
			\label{tab:non-imp-wave3}
		\end{table}

		Reading the results from Table~\ref{tab:non-imp-wave3} we determine the most promising set of 3N interaction as $V_{c_{E}}^{(0)}$, $V_{c_{D}}^{(0)}$, $V_{c_{E}}^{(1)}$, $V_{c_{D}}^{(1)}$, $V_{c_{E}}^{(l)}$ and $V_{c_{E}}^{(t)}$, and we define new non-implausible volume for inputs denoted as $\mathcal{Q}_{3}$.
		
		\item[WAVE 4\label{itm:wave-4}]  We show the results of the fitted (predicted) observables at \ref{itm:wave-3} in Fig.~\ref{fig:HM-waves-Epred} by green open (filled) right-triangle points. At this wave we use the same 3N interactions used in \ref{itm:wave-3} we run the the implausibility measure over outputs $\mathcal{Z}_{4} =$ $\mathcal{Z}_{3}$ $\, \bigcup \, $ $\{{}^{7}{\rm Li}$, ${}^{9}{\rm Be}$, ${}^{10}{\rm Be}$, ${}^{10}{\rm B}$, ${}^{11}{\rm B}$, ${}^{14}{\rm N}$, ${}^{15}{\rm N}\}$ where we included odd-odd nuclei with $N=Z$ as well as neutron-rich nuclei with $Z\le8$. Therefore, we define new non-implausible volume for inputs denoted as $\mathcal{Q}_{4}$. The results of the fitted (predicted) observables are shown by blue open (filled) circle points in Fig.~\ref{fig:HM-waves-Epred}.
		
		\item[WAVE 5\label{itm:wave-5}]
		The results in Fig.~\ref{fig:HM-waves-Epred} show that the agreements between the calculated and experimental data for light nuclei are already in an acceptable level. Therefore, by taking into consideration the level of agreement for medium-mass nuclei and the fact that we are left with two more additional 3N interactions to be used as a calibrator, at this iteration we perform a numerical experiment to determine whether these additional 3N interactions have any considerable influence on improving the agreement.  To stabilize the calculated results for medium-mass nuclei at this wave we run the implausibility measure over outputs $\mathcal{Z}_{5}=\mathcal{Z}_{4} \, \bigcup$ \, $\{{}^{40}{\rm Ca}\}$ and run the emulator for all possible sets of 3N interactions to compute the first three non-zero maximum implausibility measures. The results are shown in Table~\ref{tab:non-imp-wave5}.
		\begin{table}[ht]
			\centering
			\caption{Results for the first three non-zero maximum implausibility measures at \ref{itm:wave-5}. We run the implausibility measure over outputs ${}^3{\rm H},{}^4{\rm He},{}^7{\rm Li},{}^{8}{\rm Be},{}^{9}{\rm Be},{}^{10}{\rm Be},{}^{10}{\rm B},{}^{11}{\rm B},{}^{12}{\rm C},{}^{13}{\rm C},{}^{14}{\rm C},{}^{14}{\rm N},{}^{15}{\rm N},{}^{16}{\rm O},{}^{17}{\rm O},{}^{18}{\rm O}$, and ${}^{40}{\rm Ca}$. }
			\begin{tabular}{ |L{6.5cm} || C{2cm} | C{2cm} | C{2cm} | }
				\hline\hline
				3N interactions  &  \multicolumn{3}{c|}{ratio of non-implausible input volumes $\mathcal{Q}_{5}/\mathcal{Q}_{4}$ (\%)} \\ 
				in addition to $V_{c_{E}}^{(0)}$, $V_{c_{D}}^{(0)}$, $V_{c_{E}}^{(1)}$, $V_{c_{D}}^{(1)}$, $V_{c_{E}}^{(l)}$, $V_{c_{E}}^{(t)}$& $I^{(5)}_{M}(\mathbf{x})$ &  $I^{(6)}_{M}(\mathbf{x})$ &  $I^{(7)}_{M}(\mathbf{x})$ \\
				\hline\hline
				none & $<1$ & 64 &97\\\hline
				$V_{c_{E}}^{(2)}$		&  ~~~~2   & 23 &74\\\hline
				$V_{c_{D}}^{(2)}$		& $<1$ & 16 &97\\\hline
				$V_{c_{E}}^{(2)}$, $V_{c_{D}}^{(2)}$ &~~13  & 57   &91\\\hline\hline
			\end{tabular}
			\label{tab:non-imp-wave5}
		\end{table}
		
		Based on the numerical results given in Table~\ref{tab:non-imp-wave5}, at this iteration we run the implausibility measure over outputs $\mathcal{Z}_{5}$ using the 3N interactions $V_{c_{E}}^{(0)}$, $V_{c_{D}}^{(0)}$, $V_{c_{E}}^{(1)}$, $V_{c_{D}}^{(1)}$, $V_{c_{E}}^{(l)}$, $V_{c_{E}}^{(t)}$, $V_{c_{E}}^{(2)}$ and $V_{c_{D}}^{(2)}$ as calibrators.  The results of the fitted (predicted) observables are shown by red open (filled) square points in Fig.~\ref{fig:HM-waves-Epred}.
	\end{description}

	After performing five waves described above we discard implausible regions from the parameter spaces of the 2N LECs and we determine the minimum number of 3N interaction needed to get a descent agreement between the calculated and experimental data for the nuclear binding energies of nuclei with $A\le40$. Finally we use the final parameter spaces and perform bootstrap re-sampling of the calculated results to quantify the systematic theoretical uncertainties due to the 2N LECs.

	In the next step, using the calculated results with quantified uncertainties we perform a second analysis for the 3N interactions to show the importance of regulators and the connection with cluster EFT as well as nuclear binding energies for nuclei as discussed in the main text. To end this, we use the central values of the 2N LECs, which give the scattering phase shifts shown by blue dashed-lines in Fig.~\ref{fig:np-phase-shifts-w-errors}, with the quantified uncertainties, then we fine-tune the 3N interactions using some selected nuclear binding energies (to be explained below). In this analysis we calculate the RMSD (root
	mean square deviation) over all calculated nuclear binding energies shown in Fig.~\ref{fig:HM-waves-Epred}, and we use these results to assess the extent to which set of 3N interactions is the best for accurately describing some certain nuclear binding energies. The results are shown in Table~\ref{tab:RMSD-Step1}-\ref{tab:RMSD-Step7} where we also present the energy differences $E_{8}^{4}-2E_{4}^{2}$, $E_{12,0^{+}_{1}}^{6}-3E_{4}^{2}$,  $E_{12,{0^{+}_{2}}}^{6}-3E_{4}^{2}$, $E_{16}^{8}-4E_{4}^{2}$, $E_{6}^{2}-E_{4}^{2}$, $E_{9}^{4}-E_{8}^{4}$, $E_{13}^{6}-E_{12}^{6}$, $E_{17}^{8}-E_{16}^{8}$, $E_{10}^{4}-E_{8}^{4}$, $E_{14}^{6}-E_{12}^{6}$, and $E_{18}^{8}-E_{16}^{8}$.

	From the results shown in Tables~\ref{tab:non-imp-wave2}, \ref{tab:non-imp-wave3} and \ref{tab:non-imp-wave5}, it can be seen that nuclear binding energies are highly sensitive to the range of the three-nucleon interactions. The first few rows of Table~\ref{tab:non-imp-wave2} clearly indicates that the nearest-neighbor smearing and the next-to-nearest-neighbor smearing are playing significantly different roles in describing the light and $\alpha$-like nuclei. Therefore, we start the RMSD analysis using the simpliest two 3N interactions, $V_{c_{E}}^{(0)}$ and $V_{c_{D}}^{(0)}$, and the nuclear binding energies for light nuclei whose structure are in the form of distinct geometrical configuration of $\alpha$ clusters. The results are shown in Table~\ref{tab:RMSD-Step1}.
	\begin{table}[ht]
		\centering
		\caption{Results for the RMSD (root mean square deviation) over all calculated nuclear binding energies by using the simplest 3N interactions $V_{c_{E}}^{(0)}$ and $V_{c_{D}}^{(0)}$ to fit the nuclear binding energies for ${}^4{\rm He}$,${}^{8}{\rm Be}$,${}^{12}{\rm C}$, and ${}^{16}{\rm O}$. All energies are measured in MeV.}
		\small
		\begin{tabular}{ |l ||r |r | r | r || r | r | r | r |  }
			\hline\hline 
			3N interactions &
			RMSD    &
			$E_{8}^{4}-2E_{4}^{2}$ &  
			$E_{12,0^{+}_{1}}^{6}-3E_{4}^{2}$ &   
			$E_{16}^{8}-4E_{4}^{2}$ &   
			$E_{6}^{2}-E_{4}^{2}$ &  
			$E_{9}^{4}-E_{8}^{4}$ &    
			$E_{13}^{6}-E_{12}^{6}$ &   
			$E_{17}^{8}-E_{16}^{8}$  \\
			in addition to &  $B/A$   &          &  
			$E_{12,{0^{+}_{2}}}^{6}-3E_{4}^{2}$ &    
			&		     &  
			$E_{10}^{4}-E_{8}^{4}$ &   
			$E_{14}^{6}-E_{12}^{6}$ &  
			$E_{18}^{8}-E_{16}^{8}$   \\
			$V_{c_{E}}^{(0)}$, $V_{c_{D}}^{(0)}$  &   &
			$0.10$  & $-7.26$  & $-14.42$  &  $-0.97$  & $-1.67$   & $-4.95$ & $-4.14$     \\
			&           &              & $0.39$      &  &   &   $-8.47$   &   $-13.12$  &  $-12.19$ 		\\
			\hline\hline
			none &  $1.236$  & $-0.42(50)$	 & $-0.86(82)$	 & $7.64(76)$	 & $0.23(19)$	 & $1.47(48)$	 & $4.98(99)$	 & $-0.32(58)$	\\ 
			&            &           	 & $-0.19(92)$	 &           	 &           	 & $-1.04(69)$	 & $1.33(86)$	 & $-4.34(59)$	\\\hline  \hline
		\end{tabular}
		\label{tab:RMSD-Step1}
	\end{table}

	As seen in Table~\ref{tab:RMSD-Step1}, the RMSD of the binding energies per nucleon over all calculated nuclear binding energies is $1.236$, and it corresponds to $17.3~\%$ of the average nuclear binding energy per nucleon over all calculated nuclear binding energies $\frac{1}{N}\sum_{i = 1}^{N} z_{i}/A_{i}=7.01$, where $N$ is the number of nuclei, $z_{i}$ is the binding energy of the $i^{\rm th}$ nucleus and $A_{i}$ is the number of nucleon of the $i^{\rm th}$ nucleus.
	\begin{table}[ht]
		\centering
		\caption{Results for the RMSD over all calculated nuclear binding energies by using one additional 3N interaction as well as $V_{c_{E}}^{(0)}$ and $V_{c_{D}}^{(0)}$ to fit the nuclear binding energies for ${}^4{\rm He}$,${}^{8}{\rm Be}$,${}^{12}{\rm C}$, and ${}^{16}{\rm O}$. All energies are measured in MeV.}
		\small
		\begin{tabular}{ |l ||r |r | r | r || r | r | r | r |  }
			\hline\hline 
			3N interactions &
			RMSD    &
			$E_{8}^{4}-2E_{4}^{2}$ &  
			$E_{12,0^{+}_{1}}^{6}-3E_{4}^{2}$ &   
			$E_{16}^{8}-4E_{4}^{2}$ &   
			$E_{6}^{2}-E_{4}^{2}$ &  
			$E_{9}^{4}-E_{8}^{4}$ &    
			$E_{13}^{6}-E_{12}^{6}$ &   
			$E_{17}^{8}-E_{16}^{8}$  \\
			in addition to &  $B/A$   &          &  
			$E_{12,{0^{+}_{2}}}^{6}-3E_{4}^{2}$ &    
			&		     &  
			$E_{10}^{4}-E_{8}^{4}$ &   
			$E_{14}^{6}-E_{12}^{6}$ &  
			$E_{18}^{8}-E_{16}^{8}$   \\
			$V_{c_{E}}^{(0)}$, $V_{c_{D}}^{(0)}$  &   &
			$0.10$  & $-7.26$  & $-14.42$  &  $-0.97$  & $-1.67$   & $-4.95$ & $-4.14$     \\
			&           &              & $0.39$      &  &   &   $-8.47$   &   $-13.12$  &  $-12.19$ 		\\
			\hline\hline
			$V_{c_{E}}^{(l)}$  &  $0.403$  & $-0.57(51)$	 & $-6.16(82)$	 & $-15.17(77)$	 & $-0.17(19)$	 & $0.34(50)$	 & $0.85(98)$	 & $-1.88(59)$	\\ 
			&              &           	 & $-1.15(97)$	 &           	 &           	 & $-3.23(71)$	 & $-5.97(86)$	 & $-7.10(60)$	\\\hline 
			$V_{c_{E}}^{(2)}$  &  $0.410$  & $-0.53(52)$	 & $-5.58(82)$	 & $-15.07(76)$	 & $-0.09(19)$	 & $0.46(51)$	 & $0.78(98)$	 & $-1.86(58)$	\\ 
			&              &           	 & $-0.46(109)$	 &           	 &           	 & $-3.01(68)$	 & $-6.19(86)$	 & $-6.95(59)$	\\\hline 
			$V_{c_{E}}^{(t)}$   &  $0.415$  & $-0.30(50)$	 & $-4.04(82)$	 & $-14.45(77)$	 & $0.07(19)$	 & $0.69(50)$	 & $0.34(98)$	 & $-1.85(59)$	\\ 
			&              &           	 & $0.45(96)$	 &           	 &           	 & $-2.49(69)$	 & $-6.65(86)$	 & $-6.72(60)$	\\\hline 
			$V_{c_{D}}^{(2)}$  &  $0.419$  & $-0.30(56)$	 & $-2.28(82)$	 & $-13.05(76)$	 & $-0.12(19)$	 & $0.65(52)$	 & $-1.97(98)$	 & $-1.95(58)$	\\ 
			&              &           	 & $3.20(121)$	 &           	 &           	 & $-3.19(59)$	 & $-8.90(85)$	 & $-6.73(59)$	\\\hline 
			$V_{c_{E}}^{(1)}$  &  $0.465$  & $-0.72(57)$	 & $-5.36(82)$	 & $-14.96(76)$	 & $-0.18(19)$	 & $0.46(54)$	 & $1.66(98)$	 & $-1.62(58)$	\\ 
			&              &           	 & $-0.37(100)$	 &           	 &           	 & $-3.25(60)$	 & $-5.38(89)$	 & $-6.72(59)$	\\\hline 
			$V_{c_{D}}^{(1)}$  &  $0.542$  & $-0.09(63)$	 & $2.06(82)$	 & $-7.17(76)$	 & $0.01(19)$	 & $1.22(63)$	 & $-2.49(98)$	 & $-1.82(58)$	\\ 
			&              &           	 & $7.30(175)$	 &           	 &           	 & $-2.84(87)$	 & $-8.81(86)$	 & $-6.29(59)$	\\\hline   \hline
		\end{tabular}
		\label{tab:RMSD-Step2}
	\end{table}

	Now we use one more 3N interaction in addition to $V_{c_{E}}^{(0)}$ and $V_{c_{D}}^{(0)}$ and fit the nuclear binding energies for ${}^4{\rm He}$,${}^{8}{\rm Be}$,${}^{12}{\rm C}$, and ${}^{16}{\rm O}$. The results are given in Table~\ref{tab:RMSD-Step2}. We compute the RMSD over all calculated nuclear binding energies, and we find the lowest value as $0.403$~MeV which corresponds to $5.6~\%$ of the average nuclear binding energy per nucleon and is obtained using the 3N interactions $V_{c_{E}}^{(0)}$, $V_{c_{D}}^{(0)}$ and $V_{c_{E}}^{(l)}$.

	\begin{table}[ht]
		\centering
		\caption{Results for the RMSD over all calculated nuclear binding energies by using two additional 3N interactions as well as $V_{c_{E}}^{(0)}$ and $V_{c_{D}}^{(0)}$ to fit the nuclear binding energies for ${}^4{\rm He}$,${}^{8}{\rm Be}$,${}^{12}{\rm C}$, and ${}^{16}{\rm O}$. All energies are measured in MeV.}
		\small
		\begin{tabular}{ |l ||r |r | r | r || r | r | r | r |  }
			\hline\hline 
			3N interactions &
			RMSD    &
			$E_{8}^{4}-2E_{4}^{2}$ &  
			$E_{12,0^{+}_{1}}^{6}-3E_{4}^{2}$ &   
			$E_{16}^{8}-4E_{4}^{2}$ &   
			$E_{6}^{2}-E_{4}^{2}$ &  
			$E_{9}^{4}-E_{8}^{4}$ &    
			$E_{13}^{6}-E_{12}^{6}$ &   
			$E_{17}^{8}-E_{16}^{8}$  \\
			in addition to &  $B/A$   &          &  
			$E_{12,{0^{+}_{2}}}^{6}-3E_{4}^{2}$ &    
			&		     &  
			$E_{10}^{4}-E_{8}^{4}$ &   
			$E_{14}^{6}-E_{12}^{6}$ &  
			$E_{18}^{8}-E_{16}^{8}$   \\
			$V_{c_{E}}^{(0)}$, $V_{c_{D}}^{(0)}$  &   &
			$0.10$  & $-7.26$  & $-14.42$  &  $-0.97$  & $-1.67$   & $-4.95$ & $-4.14$     \\
			&           &              & $0.39$      &  &   &   $-8.47$   &   $-13.12$  &  $-12.19$ 		\\
			\hline\hline
			$V_{c_{E}}^{(l)}$, $V_{c_{E}}^{(t)}$  &  $0.293$  & $-0.75(49)$	 & $-7.66(85)$	 & $-15.25(76)$	 & $-0.89(19)$	 & $-0.56(48)$	 & $-1.83(100)$	 & $-2.50(58)$	\\ 
			&              &           	 & $-1.09(90)$	 &           	 &           	 & $-6.14(71)$	 & $-8.19(88)$	 & $-8.66(59)$	\\\hline 
			$V_{c_{E}}^{(2)}$, $V_{c_{E}}^{(l)}$  &  $0.342$  & $-0.29(60)$	 & $-7.49(84)$	 & $-14.80(81)$	 & $-0.92(19)$	 & $-0.76(58)$	 & $-2.19(100)$	 & $-2.56(67)$	\\ 
			&              &           	 & $-4.05(283)$	 &           	 &           	 & $-6.08(66)$	 & $-7.63(89)$	 & $-8.80(66)$	\\\hline 
			$V_{c_{D}}^{(2)}$, $V_{c_{E}}^{(l)}$  &  $0.362$  & $-0.25(58)$	 & $-7.47(82)$	 & $-14.76(78)$	 & $-0.08(19)$	 & $0.15(55)$	 & $2.57(98)$	 & $-2.13(60)$	\\ 
			&              &           	 & $-4.34(155)$	 &           	 &           	 & $-3.36(65)$	 & $-3.83(86)$	 & $-7.86(61)$	\\\hline 
			$V_{c_{D}}^{(1)}$, $V_{c_{E}}^{(l)}$  &  $0.384$  & $-0.20(59)$	 & $-7.40(82)$	 & $-14.72(77)$	 & $-0.11(19)$	 & $0.05(59)$	 & $1.05(98)$	 & $-2.03(59)$	\\ 
			&              &           	 & $-4.37(187)$	 &           	 &           	 & $-3.25(87)$	 & $-5.52(86)$	 & $-7.41(60)$	\\\hline 
			$V_{c_{D}}^{(1)}$, $V_{c_{D}}^{(2)}$  &  $0.405$  & $-0.24(49)$	 & $-4.34(82)$	 & $-14.29(76)$	 & $-0.14(19)$	 & $0.32(48)$	 & $-1.87(99)$	 & $-1.96(58)$	\\ 
			&              &           	 & $0.27(95)$	 &           	 &           	 & $-3.23(72)$	 & $-8.84(85)$	 & $-6.80(59)$	\\\hline 
			$V_{c_{E}}^{(2)}$, $V_{c_{E}}^{(t)}$   &  $0.418$  & $-0.89(54)$	 & $-7.66(82)$	 & $-15.38(76)$	 & $-0.34(19)$	 & $0.15(52)$	 & $1.33(98)$	 & $-1.87(58)$	\\ 
			&              &           	 & $-1.49(134)$	 &           	 &           	 & $-3.79(62)$	 & $-5.54(85)$	 & $-7.32(59)$	\\\hline 
			$V_{c_{E}}^{(2)}$, $V_{c_{D}}^{(1)}$  &  $0.423$  & $-0.29(54)$	 & $-7.45(82)$	 & $-14.80(77)$	 & $0.02(19)$	 & $0.23(56)$	 & $1.85(98)$	 & $-1.87(59)$	\\ 
			&              &           	 & $-4.14(139)$	 &           	 &           	 & $-2.72(90)$	 & $-4.95(86)$	 & $-7.08(60)$	\\\hline 
			$V_{c_{E}}^{(1)}$, $V_{c_{E}}^{(2)}$  &  $0.423$  & $-0.60(53)$	 & $-5.73(82)$	 & $-15.04(76)$	 & $-0.10(19)$	 & $0.47(51)$	 & $1.08(98)$	 & $-1.81(58)$	\\ 
			&              &           	 & $-0.58(108)$	 &           	 &           	 & $-2.99(68)$	 & $-5.89(86)$	 & $-6.91(59)$	\\\hline 
			$V_{c_{E}}^{(2)}$, $V_{c_{D}}^{(2)}$  &  $0.423$  & $-0.23(57)$	 & $-7.46(82)$	 & $-14.74(77)$	 & $0.24(19)$	 & $0.48(54)$	 & $4.38(98)$	 & $-1.97(59)$	\\ 
			&              &           	 & $-4.48(116)$	 &           	 &           	 & $-2.33(64)$	 & $-2.40(85)$	 & $-7.50(60)$	\\\hline 
			$V_{c_{E}}^{(1)}$, $V_{c_{E}}^{(l)}$  &  $0.427$  & $-0.67(52)$	 & $-6.32(82)$	 & $-15.24(77)$	 & $-0.18(19)$	 & $0.36(52)$	 & $1.37(98)$	 & $-1.78(59)$	\\ 
			&              &           	 & $-1.29(99)$	 &           	 &           	 & $-3.21(69)$	 & $-5.50(86)$	 & $-6.98(60)$	\\\hline 
			$V_{c_{D}}^{(1)}$, $V_{c_{E}}^{(t)}$  &  $0.430$  & $0.10(59)$	 & $-5.04(82)$	 & $-14.18(78)$	 & $0.31(19)$	 & $0.61(64)$	 & $1.26(99)$	 & $-1.84(61)$	\\ 
			&              &           	 & $-2.90(178)$	 &           	 &           	 & $-1.79(109)$	 & $-5.61(86)$	 & $-6.65(62)$	\\\hline 
			$V_{c_{E}}^{(1)}$, $V_{c_{D}}^{(1)}$  &  $0.525$  & $-0.71(49)$	 & $-7.65(83)$	 & $-15.19(77)$	 & $-0.18(19)$	 & $0.14(47)$	 & $3.87(99)$	 & $-1.39(58)$	\\ 
			&              &           	 & $-4.51(161)$	 &           	 &           	 & $-3.26(57)$	 & $-3.13(97)$	 & $-6.65(60)$	\\\hline 
			$V_{c_{E}}^{(1)}$, $V_{c_{E}}^{(t)}$  &  $0.538$  & $-1.23(73)$	 & $-6.36(84)$	 & $-15.48(76)$	 & $-0.52(19)$	 & $0.18(71)$	 & $2.85(100)$	 & $-1.37(58)$	\\ 
			&              &           	 & $-0.66(104)$	 &           	 &           	 & $-4.40(91)$	 & $-4.29(99)$	 & $-6.77(59)$	\\\hline 
			$V_{c_{E}}^{(1)}$, $V_{c_{D}}^{(2)}$  &  $0.545$  & $-1.06(59)$	 & $-7.80(82)$	 & $-15.50(77)$	 & $-0.21(19)$	 & $0.33(57)$	 & $4.91(98)$	 & $-1.28(58)$	\\ 
			&              &           	 & $-3.45(90)$	 &           	 &           	 & $-3.21(63)$	 & $-2.05(96)$	 & $-6.60(60)$	\\\hline 
			$V_{c_{D}}^{(2)}$, $V_{c_{E}}^{(t)}$  &  $0.555$  & $0.40(77)$	 & $-5.45(84)$	 & $-13.95(84)$	 & $1.01(20)$	 & $1.28(88)$	 & $6.95(100)$	 & $-1.59(70)$	\\ 
			&              &           	 & $-6.17(225)$	 &           	 &           	 & $0.48(147)$	 & $0.23(90)$	 & $-6.37(68)$	\\\hline   \hline
		\end{tabular}
		\label{tab:RMSD-Step3}
	\end{table}

	In the next analysis, we use two additional 3N interactions and fit the nuclear binding energies for ${}^4{\rm He}$,${}^{8}{\rm Be}$,${}^{12}{\rm C}$, and ${}^{16}{\rm O}$. We compute the RMSD over all calculated nuclear binding energies and the results are given in Table~\ref{tab:RMSD-Step3}. By using $V_{c_{E}}^{(0)}$, $V_{c_{D}}^{(0)}$, $V_{c_{E}}^{(l)}$ and $V_{c_{E}}^{(t)}$ we find that the lowest value for the RMSD is $0.293$~MeV, and this means that roughly we obtain the average nuclear binding energy per nucleon of all calculated nuclei with $4.1~\%$ errors. The RMSD analysis and the results shown in Table~\ref{tab:RMSD-Step3} are consistent with results of history matching given in Table~\ref{tab:non-imp-wave2} and cyan-colored up-triangle points in FIG.~\ref{fig:HM-waves-Epred}.

	\begin{table}[ht]
		\centering
		\caption{Results for the RMSD over all calculated nuclear binding energies by using three additional 3N interactions as well as $V_{c_{E}}^{(0)}$ and $V_{c_{D}}^{(0)}$ to fit the nuclear binding energies for ${}^3{\rm H},{}^4{\rm He},{}^7{\rm Li},{}^{8}{\rm Be},{}^{9}{\rm Be},{}^{10}{\rm Be},{}^{10}{\rm B},{}^{11}{\rm B},{}^{12}{\rm C},{}^{13}{\rm C},{}^{14}{\rm C},{}^{14}{\rm N},{}^{15}{\rm N},{}^{16}{\rm O},{}^{17}{\rm O}$, and ${}^{18}{\rm O}$. All energies are measured in MeV.}
		\small
		\begin{tabular}{ |l ||r |r | r | r || r | r | r | r |  }
			\hline\hline 
			3N interactions &
			RMSD    &
			$E_{8}^{4}-2E_{4}^{2}$ &  
			$E_{12,0^{+}_{1}}^{6}-3E_{4}^{2}$ &   
			$E_{16}^{8}-4E_{4}^{2}$ &   
			$E_{6}^{2}-E_{4}^{2}$ &  
			$E_{9}^{4}-E_{8}^{4}$ &    
			$E_{13}^{6}-E_{12}^{6}$ &   
			$E_{17}^{8}-E_{16}^{8}$  \\
			in addition to &  $B/A$   &          &  
			$E_{12,{0^{+}_{2}}}^{6}-3E_{4}^{2}$ &    
			&		     &  
			$E_{10}^{4}-E_{8}^{4}$ &   
			$E_{14}^{6}-E_{12}^{6}$ &  
			$E_{18}^{8}-E_{16}^{8}$   \\
			$V_{c_{E}}^{(0)}$, $V_{c_{D}}^{(0)}$  &   &
			$0.10$  & $-7.26$  & $-14.42$  &  $-0.97$  & $-1.67$   & $-4.95$ & $-4.14$     \\
			&           &              & $0.39$      &  &   &   $-8.47$   &   $-13.12$  &  $-12.19$ 		\\
			\hline\hline
			$V_{c_{E}}^{(1)}$, $V_{c_{E}}^{(l)}$  &  $0.131$  & $-0.52(51)$	 & $-8.47(85)$	 & $-17.40(76)$	 & $-0.98(19)$	 & $-0.86(50)$	 & $-3.97(100)$	 & $-3.07(58)$	\\ 
			$V_{c_{E}}^{(t)}$  &              &           	 & $-1.30(97)$	 &           	 &           	 & $-6.69(66)$	 & $-10.38(92)$	 & $-9.56(59)$	\\\hline 
			$V_{c_{E}}^{(1)}$, $V_{c_{E}}^{(2)}$  &  $0.169$  & $0.13(61)$	 & $-5.55(82)$	 & $-16.61(78)$	 & $-0.25(19)$	 & $-0.02(61)$	 & $-3.96(98)$	 & $-2.95(61)$	\\ 
			$V_{c_{E}}^{(l)}$  &              &           	 & $-0.44(107)$	 &           	 &           	 & $-4.02(90)$	 & $-10.56(92)$	 & $-8.57(61)$	\\\hline 
			$V_{c_{E}}^{(1)}$, $V_{c_{D}}^{(2)}$  &  $0.184$  & $0.11(61)$	 & $-5.93(82)$	 & $-16.52(78)$	 & $-0.15(19)$	 & $0.07(63)$	 & $-3.47(99)$	 & $-2.92(60)$	\\ 
			$V_{c_{E}}^{(l)}$  &              &           	 & $-0.71(91)$	 &           	 &           	 & $-3.59(102)$	 & $-10.17(95)$	 & $-8.46(61)$	\\\hline 
			$V_{c_{E}}^{(1)}$, $V_{c_{D}}^{(1)}$ &  $0.188$  & $0.22(70)$	 & $-6.09(82)$	 & $-16.34(78)$	 & $-0.13(19)$	 & $0.01(72)$	 & $-3.58(99)$	 & $-2.96(61)$	\\ 
			$V_{c_{E}}^{(l)}$   &              &           	 & $-1.38(117)$	 &           	 &           	 & $-3.59(112)$	 & $-10.20(95)$	 & $-8.52(61)$	\\\hline 
			$V_{c_{E}}^{(2)}$, $V_{c_{D}}^{(1)}$ &  $0.204$  & $-0.12(55)$	 & $-5.83(83)$	 & $-17.25(78)$	 & $-0.39(19)$	 & $-0.16(52)$	 & $-2.01(99)$	 & $-2.54(61)$	\\ 
			$V_{c_{E}}^{(l)}$   &              &           	 & $-1.60(185)$	 &           	 &           	 & $-4.50(59)$	 & $-8.88(87)$	 & $-8.22(61)$	\\\hline 
			$V_{c_{E}}^{(1)}$, $V_{c_{E}}^{(2)}$  &  $0.215$  & $-1.32(53)$	 & $-11.54(82)$	 & $-18.00(77)$	 & $-0.81(19)$	 & $-0.59(49)$	 & $0.24(98)$	 & $-2.43(61)$	\\ 
			$V_{c_{E}}^{(t)}$  &              &           	 & $-2.80(202)$	 &           	 &           	 & $-5.52(55)$	 & $-6.86(90)$	 & $-8.71(61)$	\\\hline 
			$V_{c_{E}}^{(2)}$, $V_{c_{E}}^{(l)}$ &  $0.217$  & $-1.21(51)$	 & $-10.95(85)$	 & $-18.60(77)$	 & $-1.20(19)$	 & $-1.05(52)$	 & $-1.30(101)$	 & $-2.65(59)$	\\ 
			$V_{c_{E}}^{(t)}$   &              &           	 & $-2.77(95)$	 &           	 &           	 & $-7.18(80)$	 & $-8.02(89)$	 & $-9.35(60)$	\\\hline 
			$V_{c_{E}}^{(2)}$, $V_{c_{D}}^{(2)}$ &  $0.230$  & $-0.58(53)$	 & $-5.69(82)$	 & $-18.51(77)$	 & $-0.60(19)$	 & $-0.13(50)$	 & $-2.95(99)$	 & $-2.36(60)$	\\ 
			$V_{c_{E}}^{(l)}$ &              &           	 & $0.51(89)$	 &           	 &           	 & $-4.78(62)$	 & $-10.06(86)$	 & $-7.84(60)$	\\\hline 
			$V_{c_{D}}^{(1)}$, $V_{c_{E}}^{(l)}$ &  $0.236$  & $-1.42(70)$	 & $-9.45(86)$	 & $-19.01(77)$	 & $-1.41(19)$	 & $-0.97(74)$	 & $-2.68(102)$	 & $-2.72(59)$	\\ 
			$V_{c_{E}}^{(t)}$   &              &           	 & $0.47(194)$	 &           	 &           	 & $-7.81(119)$	 & $-9.40(90)$	 & $-9.44(60)$	\\\hline 
			$V_{c_{E}}^{(1)}$, $V_{c_{E}}^{(2)}$  &  $0.240$  & $-0.54(52)$	 & $-5.42(82)$	 & $-17.28(76)$	 & $-0.21(19)$	 & $0.31(50)$	 & $-1.09(98)$	 & $-2.17(58)$	\\ 
			$V_{c_{D}}^{(2)}$  &              &           	 & $0.56(122)$	 &           	 &           	 & $-3.46(64)$	 & $-8.53(85)$	 & $-7.37(59)$	\\\hline 
			$V_{c_{D}}^{(1)}$, $V_{c_{D}}^{(2)}$ &  $0.249$  & $-0.40(52)$	 & $-7.48(82)$	 & $-17.86(77)$	 & $-0.26(19)$	 & $-0.08(52)$	 & $-0.38(99)$	 & $-2.18(59)$	\\ 
			$V_{c_{E}}^{(l)}$   &              &           	 & $-3.13(134)$	 &           	 &           	 & $-3.69(80)$	 & $-7.67(86)$	 & $-7.58(60)$	\\\hline 
			$V_{c_{E}}^{(1)}$, $V_{c_{E}}^{(2)}$ &  $0.250$  & $-0.28(55)$	 & $-7.37(82)$	 & $-16.83(77)$	 & $-0.03(19)$	 & $0.16(56)$	 & $0.11(99)$	 & $-2.29(59)$	\\ 
			$V_{c_{D}}^{(1)}$   &              &           	 & $-2.96(98)$	 &           	 &           	 & $-2.98(92)$	 & $-7.08(86)$	 & $-7.65(60)$	\\\hline 
			$V_{c_{E}}^{(2)}$, $V_{c_{D}}^{(1)}$  &  $0.252$  & $-0.43(49)$	 & $-5.99(82)$	 & $-17.49(76)$	 & $-0.26(19)$	 & $0.08(49)$	 & $-1.66(99)$	 & $-2.13(58)$	\\ 
			$V_{c_{D}}^{(2)}$   &              &           	 & $-0.92(96)$	 &           	 &           	 & $-3.60(76)$	 & $-9.14(85)$	 & $-7.25(59)$	\\\hline 
			$V_{c_{D}}^{(2)}$, $V_{c_{E}}^{(l)}$  &  $0.255$  & $-1.24(62)$	 & $-8.95(88)$	 & $-18.75(77)$	 & $-1.47(19)$	 & $-1.14(67)$	 & $-4.83(103)$	 & $-2.82(60)$	\\ 
			$V_{c_{E}}^{(t)}$ &              &           	 & $0.73(128)$	 &           	 &           	 & $-8.11(109)$	 & $-11.61(91)$	 & $-9.37(60)$	\\\hline 
			$V_{c_{E}}^{(1)}$, $V_{c_{D}}^{(1)}$  &  $0.263$  & $-0.51(49)$	 & $-6.28(82)$	 & $-17.75(76)$	 & $-0.29(19)$	 & $0.05(49)$	 & $-1.27(99)$	 & $-2.06(58)$	\\ 
			$V_{c_{D}}^{(2)}$ &              &           	 & $-1.21(95)$	 &           	 &           	 & $-3.66(73)$	 & $-8.78(86)$	 & $-7.21(59)$	\\\hline 
			$V_{c_{E}}^{(2)}$, $V_{c_{D}}^{(2)}$ &  $0.264$  & $-1.54(70)$	 & $-10.22(83)$	 & $-18.31(79)$	 & $-1.07(19)$	 & $-0.67(71)$	 & $-0.88(99)$	 & $-2.31(63)$	\\ 
			$V_{c_{E}}^{(t)}$   &              &           	 & $-0.60(232)$	 &           	 &           	 & $-6.42(93)$	 & $-8.19(88)$	 & $-8.54(62)$	\\\hline 
			$V_{c_{E}}^{(2)}$, $V_{c_{D}}^{(1)}$ &  $0.265$  & $-1.65(66)$	 & $-11.56(83)$	 & $-18.64(78)$	 & $-0.99(19)$	 & $-0.65(65)$	 & $1.05(99)$	 & $-2.20(62)$	\\ 
			$V_{c_{E}}^{(t)}$   &              &           	 & $-2.36(229)$	 &           	 &           	 & $-6.08(81)$	 & $-6.21(87)$	 & $-8.56(62)$	\\\hline 
			$V_{c_{D}}^{(1)}$, $V_{c_{D}}^{(2)}$  &  $0.267$  & $-0.63(58)$	 & $-4.90(82)$	 & $-17.43(76)$	 & $-0.59(19)$	 & $-0.09(55)$	 & $-4.51(99)$	 & $-2.23(58)$	\\ 
			$V_{c_{E}}^{(t)}$  &              &           	 & $2.13(108)$	 &           	 &           	 & $-4.66(64)$	 & $-12.07(85)$	 & $-7.31(59)$	\\\hline 
			$V_{c_{E}}^{(1)}$, $V_{c_{D}}^{(2)}$   &  $0.320$  & $-1.00(75)$	 & $-4.62(85)$	 & $-17.46(76)$	 & $-0.64(19)$	 & $0.07(75)$	 & $-1.05(100)$	 & $-1.90(58)$	\\ 
			$V_{c_{E}}^{(t)}$  &              &           	 & $2.40(132)$	 &           	 &           	 & $-4.98(103)$	 & $-8.75(93)$	 & $-7.28(59)$	\\\hline 
			$V_{c_{E}}^{(1)}$, $V_{c_{D}}^{(1)}$  &  $0.343$  & $-1.16(63)$	 & $-7.97(82)$	 & $-18.56(76)$	 & $-0.47(19)$	 & $-0.00(60)$	 & $2.78(98)$	 & $-1.61(58)$	\\ 
			$V_{c_{E}}^{(t)}$   &              &           	 & $-2.36(98)$	 &           	 &           	 & $-4.16(65)$	 & $-4.84(94)$	 & $-7.15(59)$	\\\hline \hline
		\end{tabular}
		\label{tab:RMSD-Step4}
	\end{table}

	After finding four 3N interactions which give a good description for light $\alpha$-like nuclei, we include neutron-rich nuclei in RMSD analyses. We use three 3N interactions in addition to  $V_{c_{E}}^{(0)}$ and $V_{c_{D}}^{(0)}$ and fit the nuclear binding energies for ${}^3{\rm H},{}^4{\rm He},{}^7{\rm Li},{}^{8}{\rm Be},{}^{9}{\rm Be},{}^{10}{\rm Be},{}^{10}{\rm B},{}^{11}{\rm B},{}^{12}{\rm C},{}^{13}{\rm C},{}^{14}{\rm C},{}^{14}{\rm N},{}^{15}{\rm N},{}^{16}{\rm O},{}^{17}{\rm O}$, and ${}^{18}{\rm O}$. We compute the RMSD over all calculated nuclear binding energies and the results are given in Table~\ref{tab:RMSD-Step4}. We find that the lowest value for the RMSD  is $0.131$~MeV and is obtained by using $V_{c_{E}}^{(0)}$, $V_{c_{D}}^{(0)}$, $V_{c_{E}}^{(l)}$, $V_{c_{E}}^{(t)}$, and $V_{c_{E}}^{(1)}$, which means that the average nuclear binding energy per nucleon of all calculated nuclei is computed roughly with $1.83~\%$ errors. The RMSD analysis and the results shown in Table~\ref{tab:RMSD-Step4}.

	\begin{table}[ht]
		\centering\caption{Results for the RMSD over all calculated nuclear binding energies by using four additional 3N interactions as well as $V_{c_{E}}^{(0)}$ and $V_{c_{D}}^{(0)}$ to fit the nuclear binding energies for ${}^3{\rm H},{}^4{\rm He},{}^7{\rm Li},{}^{8}{\rm Be},{}^{9}{\rm Be},{}^{10}{\rm Be},{}^{10}{\rm B},{}^{11}{\rm B},{}^{12}{\rm C},{}^{13}{\rm C},{}^{14}{\rm C},{}^{14}{\rm N},{}^{15}{\rm N},{}^{16}{\rm O},{}^{17}{\rm O}$, and ${}^{18}{\rm O}$. All energies are measured in MeV.}
		\small
		\begin{tabular}{ |l ||r |r | r | r || r | r | r | r |  }
			\hline\hline 
			3N interactions &
			RMSD    &
			$E_{8}^{4}-2E_{4}^{2}$ &  
			$E_{12,0^{+}_{1}}^{6}-3E_{4}^{2}$ &   
			$E_{16}^{8}-4E_{4}^{2}$ &   
			$E_{6}^{2}-E_{4}^{2}$ &  
			$E_{9}^{4}-E_{8}^{4}$ &    
			$E_{13}^{6}-E_{12}^{6}$ &   
			$E_{17}^{8}-E_{16}^{8}$  \\
			in addition to &  $B/A$   &          &  
			$E_{12,{0^{+}_{2}}}^{6}-3E_{4}^{2}$ &    
			&		     &  
			$E_{10}^{4}-E_{8}^{4}$ &   
			$E_{14}^{6}-E_{12}^{6}$ &  
			$E_{18}^{8}-E_{16}^{8}$   \\
			$V_{c_{E}}^{(0)}$, $V_{c_{D}}^{(0)}$  &   &
			$0.10$  & $-7.26$  & $-14.42$  &  $-0.97$  & $-1.67$   & $-4.95$ & $-4.14$     \\
			&           &              & $0.39$      &  &   &   $-8.47$   &   $-13.12$  &  $-12.19$ 		\\
			\hline\hline
			$V_{c_{E}}^{(1)}$, $V_{c_{E}}^{(2)}$  &  $0.109$  & $-0.67(51)$	 & $-9.63(84)$	 & $-17.18(77)$	 & $-1.08(19)$	 & $-1.00(51)$	 & $-4.27(100)$	 & $-3.22(59)$	\\ 
			$V_{c_{E}}^{(l)}$, $V_{c_{E}}^{(t)}$  &              &           	 & $-1.34(92)$	 &           	 &           	 & $-7.05(70)$	 & $-10.64(97)$	 & $-9.91(60)$	\\\hline 
			$V_{c_{E}}^{(1)}$, $V_{c_{D}}^{(1)}$    &  $0.159$  & $-0.98(55)$	 & $-8.13(84)$	 & $-17.98(76)$	 & $-1.27(19)$	 & $-0.88(58)$	 & $-4.83(100)$	 & $-3.11(58)$	\\ 
			$V_{c_{E}}^{(l)}$, $V_{c_{E}}^{(t)}$  &              &           	 & $1.58(163)$	 &           	 &           	 & $-7.59(93)$	 & $-11.30(92)$	 & $-9.75(59)$	\\\hline 
			$V_{c_{E}}^{(1)}$, $V_{c_{E}}^{(2)}$   &  $0.165$  & $-0.13(51)$	 & $-5.61(82)$	 & $-17.34(78)$	 & $-0.38(19)$	 & $-0.06(52)$	 & $-4.11(99)$	 & $-2.78(60)$	\\ 
			$V_{c_{D}}^{(2)}$, $V_{c_{E}}^{(l)}$  &              &           	 & $0.18(89)$	 &           	 &           	 & $-4.26(84)$	 & $-10.87(89)$	 & $-8.29(61)$	\\\hline 
			$V_{c_{E}}^{(1)}$, $V_{c_{E}}^{(2)}$    &  $0.169$  & $0.13(62)$	 & $-5.53(82)$	 & $-16.60(78)$	 & $-0.25(19)$	 & $-0.03(61)$	 & $-3.95(98)$	 & $-2.95(61)$	\\ 
			$V_{c_{D}}^{(1)}$, $V_{c_{E}}^{(l)}$   &              &           	 & $-0.47(109)$	 &           	 &           	 & $-4.03(89)$	 & $-10.55(92)$	 & $-8.56(61)$	\\\hline 
			$V_{c_{E}}^{(1)}$, $V_{c_{D}}^{(1)}$   &  $0.185$  & $0.28(74)$	 & $-6.09(82)$	 & $-16.35(78)$	 & $-0.14(19)$	 & $-0.06(76)$	 & $-3.84(99)$	 & $-2.91(60)$	\\ 
			$V_{c_{D}}^{(2)}$, $V_{c_{E}}^{(l)}$  &              &           	 & $-1.80(153)$	 &           	 &           	 & $-3.58(118)$	 & $-10.57(93)$	 & $-8.38(61)$	\\\hline 
			$V_{c_{E}}^{(1)}$, $V_{c_{E}}^{(2)}$  &  $0.202$  & $-1.29(53)$	 & $-11.38(82)$	 & $-17.90(77)$	 & $-0.77(19)$	 & $-0.53(49)$	 & $-0.12(98)$	 & $-2.50(60)$	\\ 
			$V_{c_{D}}^{(1)}$, $V_{c_{E}}^{(t)}$   &              &           	 & $-2.41(227)$	 &           	 &           	 & $-5.39(56)$	 & $-7.17(93)$	 & $-8.73(61)$	\\\hline 
			$V_{c_{E}}^{(2)}$, $V_{c_{D}}^{(1)}$    &  $0.214$  & $-0.19(51)$	 & $-5.64(82)$	 & $-17.82(77)$	 & $-0.52(19)$	 & $-0.28(49)$	 & $-3.76(99)$	 & $-2.47(60)$	\\ 
			$V_{c_{D}}^{(2)}$, $V_{c_{E}}^{(l)}$    &              &           	 & $-0.88(168)$	 &           	 &           	 & $-4.69(66)$	 & $-10.79(86)$	 & $-7.90(61)$	\\\hline 
			$V_{c_{E}}^{(1)}$, $V_{c_{E}}^{(2)}$  &  $0.228$  & $-1.39(62)$	 & $-10.34(82)$	 & $-18.04(78)$	 & $-0.97(19)$	 & $-0.63(61)$	 & $-1.30(98)$	 & $-2.44(62)$	\\ 
			$V_{c_{D}}^{(2)}$, $V_{c_{E}}^{(t)}$  &              &           	 & $-0.83(236)$	 &           	 &           	 & $-6.09(72)$	 & $-8.51(89)$	 & $-8.64(62)$	\\\hline 
			$V_{c_{E}}^{(1)}$, $V_{c_{D}}^{(2)}$  &  $0.236$  & $-1.12(58)$	 & $-8.74(87)$	 & $-18.52(77)$	 & $-1.43(19)$	 & $-1.12(63)$	 & $-5.11(102)$	 & $-2.89(59)$	\\ 
			$V_{c_{E}}^{(l)}$, $V_{c_{E}}^{(t)}$   &              &           	 & $0.76(120)$	 &           	 &           	 & $-7.99(102)$	 & $-11.83(91)$	 & $-9.43(60)$	\\\hline 
			$V_{c_{E}}^{(2)}$, $V_{c_{D}}^{(1)}$   &  $0.244$  & $-1.55(72)$	 & $-10.31(86)$	 & $-19.09(77)$	 & $-1.45(19)$	 & $-1.05(77)$	 & $-2.21(102)$	 & $-2.69(60)$	\\ 
			$V_{c_{E}}^{(l)}$, $V_{c_{E}}^{(t)}$  &              &           	 & $-0.09(211)$	 &           	 &           	 & $-7.94(123)$	 & $-8.96(90)$	 & $-9.50(60)$	\\\hline 
			$V_{c_{D}}^{(1)}$, $V_{c_{D}}^{(2)}$    &  $0.245$  & $-1.02(53)$	 & $-8.98(87)$	 & $-18.50(77)$	 & $-1.39(19)$	 & $-1.18(56)$	 & $-5.03(102)$	 & $-2.83(59)$	\\ 
			$V_{c_{E}}^{(l)}$, $V_{c_{E}}^{(t)}$   &              &           	 & $-0.18(90)$	 &           	 &           	 & $-7.83(88)$	 & $-11.80(91)$	 & $-9.30(60)$	\\\hline 
			$V_{c_{E}}^{(2)}$, $V_{c_{D}}^{(2)}$  &  $0.258$  & $-1.32(64)$	 & $-9.56(88)$	 & $-18.82(78)$	 & $-1.49(19)$	 & $-1.19(69)$	 & $-4.37(103)$	 & $-2.79(61)$	\\ 
			$V_{c_{E}}^{(l)}$, $V_{c_{E}}^{(t)}$   &              &           	 & $0.31(141)$	 &           	 &           	 & $-8.15(112)$	 & $-11.17(92)$	 & $-9.41(61)$	\\\hline 
			$V_{c_{E}}^{(1)}$, $V_{c_{E}}^{(2)}$  &  $0.294$  & $-0.69(53)$	 & $-6.18(82)$	 & $-18.28(76)$	 & $-0.41(19)$	 & $0.00(52)$	 & $-1.19(99)$	 & $-1.91(58)$	\\ 
			$V_{c_{D}}^{(1)}$, $V_{c_{D}}^{(2)}$  &              &           	 & $-0.76(90)$	 &           	 &           	 & $-3.96(68)$	 & $-8.83(88)$	 & $-7.02(59)$	\\\hline 
			$V_{c_{E}}^{(2)}$, $V_{c_{D}}^{(1)}$  &  $0.298$  & $-1.46(61)$	 & $-11.06(82)$	 & $-19.05(78)$	 & $-1.09(19)$	 & $-0.88(58)$	 & $-1.80(99)$	 & $-2.22(61)$	\\ 
			$V_{c_{D}}^{(2)}$, $V_{c_{E}}^{(t)}$    &              &           	 & $-2.19(150)$	 &           	 &           	 & $-6.27(63)$	 & $-9.16(86)$	 & $-8.23(61)$	\\\hline 
			$V_{c_{E}}^{(1)}$, $V_{c_{D}}^{(1)}$ &  $0.367$  & $-1.29(74)$	 & $-7.65(85)$	 & $-18.89(77)$	 & $-0.93(19)$	 & $-0.47(72)$	 & $-1.20(101)$	 & $-1.80(59)$	\\ 
			$V_{c_{D}}^{(2)}$, $V_{c_{E}}^{(t)}$      &              &           	 & $-0.36(100)$	 &           	 &           	 & $-5.69(83)$	 & $-8.92(95)$	 & $-7.33(60)$	\\\hline  \hline
		\end{tabular}
		\label{tab:RMSD-Step5}
	\end{table}

	Now we fit the nuclear binding energies for ${}^3{\rm H},{}^4{\rm He},{}^7{\rm Li},{}^{8}{\rm Be},{}^{9}{\rm Be},{}^{10}{\rm Be},{}^{10}{\rm B},{}^{11}{\rm B},{}^{12}{\rm C},{}^{13}{\rm C},{}^{14}{\rm C},{}^{14}{\rm N},{}^{15}{\rm N},{}^{16}{\rm O},{}^{17}{\rm O}$, and ${}^{18}{\rm O}$ using four and five 3N interactions in addition to $V_{c_{E}}^{(0)}$ and $V_{c_{D}}^{(0)}$, and the results are shown in Table~\ref{tab:RMSD-Step5} and Table~\ref{tab:RMSD-Step6}, respectively. The lowest value for the RMSD is found as $0.109$($0.102$)~MeV when six(seven) 3N interactions are used in the fit, and this value corresponds to $1.54~\%$($1.43\%$) of the average nuclear binding energy per nucleon.

	\begin{table}[ht]
		\centering\caption{Results for the RMSD over all calculated nuclear binding energies by using five additional 3N interactions as well as $V_{c_{E}}^{(0)}$ and $V_{c_{D}}^{(0)}$ to fit the nuclear binding energies for ${}^3{\rm H},{}^4{\rm He},{}^7{\rm Li},{}^{8}{\rm Be},{}^{9}{\rm Be},{}^{10}{\rm Be},{}^{10}{\rm B},{}^{11}{\rm B},{}^{12}{\rm C},{}^{13}{\rm C},{}^{14}{\rm C},{}^{14}{\rm N},{}^{15}{\rm N},{}^{16}{\rm O},{}^{17}{\rm O}$, and ${}^{18}{\rm O}\}$. All energies are measured in MeV. }
		\small
		\begin{tabular}{ |l ||r |r | r | r || r | r | r | r |  }
			\hline\hline 
			3N interactions &
			RMSD    &
			$E_{8}^{4}-2E_{4}^{2}$ &  
			$E_{12,0^{+}_{1}}^{6}-3E_{4}^{2}$ &   
			$E_{16}^{8}-4E_{4}^{2}$ &   
			$E_{6}^{2}-E_{4}^{2}$ &  
			$E_{9}^{4}-E_{8}^{4}$ &    
			$E_{13}^{6}-E_{12}^{6}$ &   
			$E_{17}^{8}-E_{16}^{8}$  \\
			in addition to &  $B/A$   &          &  
			$E_{12,{0^{+}_{2}}}^{6}-3E_{4}^{2}$ &    
			&		     &  
			$E_{10}^{4}-E_{8}^{4}$ &   
			$E_{14}^{6}-E_{12}^{6}$ &  
			$E_{18}^{8}-E_{16}^{8}$   \\
			$V_{c_{E}}^{(0)}$, $V_{c_{D}}^{(0)}$  &   &
			$0.10$  & $-7.26$  & $-14.42$  &  $-0.97$  & $-1.67$   & $-4.95$ & $-4.14$     \\
			&           &              & $0.39$      &  &   &   $-8.47$   &   $-13.12$  &  $-12.19$ 		\\
			\hline\hline
			$V_{c_{E}}^{(1)}$, $V_{c_{E}}^{(2)}$, $V_{c_{D}}^{(1)}$  &  $0.102$  & $-0.73(51)$	 & $-9.99(84)$	 & $-17.60(77)$	 & $-1.11(19)$	 & $-1.05(51)$	 & $-4.15(100)$	 & $-3.24(59)$	\\ 
			$V_{c_{E}}^{(l)}$, $V_{c_{E}}^{(t)}$  &    & 	 & $-1.64(93)$	 & 	 & 	 & $-7.14(69)$	 & $-10.61(97)$	 & $-9.98(59)$	\\\hline 
			$V_{c_{E}}^{(1)}$, $V_{c_{E}}^{(2)}$, $V_{c_{D}}^{(2)}$  &  $0.144$  & $-0.91(49)$	 & $-9.93(84)$	 & $-17.80(77)$	 & $-1.20(19)$	 & $-1.06(49)$	 & $-4.44(99)$	 & $-3.10(59)$	\\ 
			$V_{c_{E}}^{(l)}$, $V_{c_{E}}^{(t)}$ &              &            	 & $-0.85(117)$	 &            	 &            	 & $-7.30(67)$	 & $-10.96(95)$	 & $-9.73(60)$	\\\hline 
			$V_{c_{E}}^{(1)}$, $V_{c_{E}}^{(2)}$, $V_{c_{D}}^{(1)}$ &  $0.185$  & $0.02(57)$	 & $-5.59(82)$	 & $-17.22(77)$	 & $-0.40(19)$	 & $-0.23(55)$	 & $-3.80(99)$	 & $-2.66(60)$	\\ 
			$V_{c_{D}}^{(2)}$, $V_{c_{E}}^{(l)}$  &              &            	 & $-1.24(179)$	 &            	 &            	 & $-4.46(72)$	 & $-10.67(86)$	 & $-8.16(61)$	\\\hline 
			$V_{c_{E}}^{(2)}$, $V_{c_{D}}^{(1)}$, $V_{c_{D}}^{(2)}$  &  $0.260$  & $-1.30(61)$	 & $-9.63(87)$	 & $-18.85(77)$	 & $-1.48(19)$	 & $-1.20(65)$	 & $-4.39(102)$	 & $-2.78(60)$	\\ 
			$V_{c_{E}}^{(l)}$, $V_{c_{E}}^{(t)}$ &              &            	 & $0.09(130)$	 &            	 &            	 & $-8.07(102)$	 & $-11.20(91)$	 & $-9.36(61)$	\\\hline 
			$V_{c_{E}}^{(1)}$, $V_{c_{D}}^{(1)}$, $V_{c_{D}}^{(2)}$  &  $0.270$  & $-1.05(52)$	 & $-9.43(87)$	 & $-18.71(77)$	 & $-1.39(19)$	 & $-1.24(53)$	 & $-4.70(102)$	 & $-2.71(59)$	\\ 
			$V_{c_{E}}^{(l)}$, $V_{c_{E}}^{(t)}$   &              &            	 & $-1.03(92)$	 &            	 &            	 & $-7.76(79)$	 & $-11.57(91)$	 & $-9.11(60)$	\\\hline 
			$V_{c_{E}}^{(1)}$, $V_{c_{E}}^{(2)}$, $V_{c_{D}}^{(1)}$  &  $0.337$  & $-1.57(69)$	 & $-10.88(84)$	 & $-19.30(78)$	 & $-1.18(19)$	 & $-0.93(66)$	 & $-1.32(99)$	 & $-2.08(62)$	\\ 
			$V_{c_{D}}^{(2)}$, $V_{c_{E}}^{(t)}$  &              &            	 & $-2.02(147)$	 &            	 &            	 & $-6.57(75)$	 & $-8.79(87)$	 & $-8.13(62)$	\\\hline   \hline
		\end{tabular}
		\label{tab:RMSD-Step6}
	\end{table}

	Finally, we fit the nuclear binding energies for ${}^3{\rm H},{}^4{\rm He},{}^7{\rm Li},{}^{8}{\rm Be},{}^{9}{\rm Be},{}^{10}{\rm Be},{}^{10}{\rm B},{}^{11}{\rm B},{}^{12}{\rm C},{}^{13}{\rm C},{}^{14}{\rm C},{}^{14}{\rm N},{}^{15}{\rm N},{}^{16}{\rm O},{}^{17}{\rm O},{}^{18}{\rm O}$, and ${}^{40}{\rm Ca}$ using six 3N interactions in addition to $V_{c_{E}}^{(0)}$ and $V_{c_{D}}^{(0)}$, and the results are shown in Table~\ref{tab:RMSD-Step7}. The lowest value for the RMSD is found as $0.079$~MeV, which corresponds to $1.11~\%$ of the average nuclear binding energy per nucleon.

	\begin{table}[ht]
		\centering
		\caption{Results for the RMSD over all calculated nuclear binding energies by using six additional 3N interactions as well as $V_{c_{E}}^{(0)}$ and $V_{c_{D}}^{(0)}$ to fit the nuclear binding energies for ${}^3{\rm H},{}^4{\rm He},{}^7{\rm Li},{}^{8}{\rm Be},{}^{9}{\rm Be},{}^{10}{\rm Be},{}^{10}{\rm B},{}^{11}{\rm B},{}^{12}{\rm C},{}^{13}{\rm C},{}^{14}{\rm C},{}^{14}{\rm N},{}^{15}{\rm N},{}^{16}{\rm O},{}^{17}{\rm O},{}^{18}{\rm O}$, and ${}^{40}{\rm Ca}$. All energies are measured in MeV.}
		\small
		\begin{tabular}{ |l ||r |r | r | r || r | r | r | r |  }
			\hline\hline 
			3N interactions &
			RMSD    &
			$E_{8}^{4}-2E_{4}^{2}$ &  
			$E_{12,0^{+}_{1}}^{6}-3E_{4}^{2}$ &   
			$E_{16}^{8}-4E_{4}^{2}$ &   
			$E_{6}^{2}-E_{4}^{2}$ &  
			$E_{9}^{4}-E_{8}^{4}$ &    
			$E_{13}^{6}-E_{12}^{6}$ &   
			$E_{17}^{8}-E_{16}^{8}$  \\
			in addition to &  $B/A$   &          &  
			$E_{12,{0^{+}_{2}}}^{6}-3E_{4}^{2}$ &    
			&		     &  
			$E_{10}^{4}-E_{8}^{4}$ &   
			$E_{14}^{6}-E_{12}^{6}$ &  
			$E_{18}^{8}-E_{16}^{8}$   \\
			$V_{c_{E}}^{(0)}$, $V_{c_{D}}^{(0)}$  &   &
			$0.10$  & $-7.26$  & $-14.42$  &  $-0.97$  & $-1.67$   & $-4.95$ & $-4.14$     \\
			&           &              & $0.39$      &  &   &   $-8.47$   &   $-13.12$  &  $-12.19$ 		\\
			\hline\hline
			$V_{c_{E}}^{(1)}$, $V_{c_{E}}^{(2)}$, $V_{c_{D}}^{(1)}$  &  $0.079$  & $-0.65(49)$	 & $-8.55(84)$	 & $-17.45(77)$	 & $-1.09(19)$	 & $-0.85(53)$	 & $-5.44(100)$	 & $-3.48(59)$	\\ 
			$V_{c_{D}}^{(2)}$, $V_{c_{E}}^{(l)}$, $V_{c_{E}}^{(t)}$    &    &  	 & $0.99(168)$	 & 	 &    	 & $-7.19(83)$	 & $-11.79(106)$	 & $-10.23(60)$	\\\hline    \hline
		\end{tabular}
		\label{tab:RMSD-Step7}
	\end{table}

	\subsection{Results}
	
	In this subsection, we give numerical details of the results presented in the main text.  We have calculated results for the ground state and excited state energies as well as the charge radii of some selected nuclei up to $A = 40$, pure neutron matter, and symmetric nuclear matter at N3LO in chiral effective field theory using wave function matching. In Table~\ref{tab:RMS-charge_radii} we present the numerical details of the results for the calculated ground state and excited state energies and charge radii as well as experimental data for comparison. The quoted errors includes all the statistic and systematic uncertainties such as computational uncertainties
	from Monte Carlo errors, infinite volume extrapolation, infinite projection time extrapolation, truncation error of the expansion of the $H'-H^{S}$ and the systematic uncertainty due to the fitting the 2N short-range interactions. The errors on the charge radii grow for the heaviest nuclei, and the main sources of these errors are Monte Carlo statistics. Therefore, they can be reduced by using larger computer resources. In Table~\ref{tab:Energy-neutron-matter} we show the calculated pure neutron matter energies for various numbers of neutrons and box sizes as well as corresponding densities. Similarly, large scale calculations corresponding to high densities have larger errors which can be reduced by larger computer resources.  In Table~\ref{tab:Energy-nuclear-matter} we show the results for symmetric nuclear matter energies for various numbers of nucleons as well as corresponding densities. The large errors on the results with large numbers of nucleon are the statistical errors of Monte Carlo calculations and can be reduced by more computer power.
	\begin{table}[ht]
		\centering
		\caption{Results of calculated ground state and excited state energies of some selected nuclei at N3LO in chiral effective field theory and comparison with experimental data.}
		\small
		\begin{tabular}{ |c || c  | c  || c  | c |  }
			\hline\hline 
			\quad~Nuclei~\quad     &
			\hspace{0.5cm}	$B$~(MeV)\hspace{0.5cm}	   &  
			\quad~Experiment~(MeV)~\quad  &
			\hspace{0.5cm}	$R_{c}$~(fm)\hspace{0.5cm}		   &  
			\quad~Experiment~(fm)~\quad		\\
			\hline\hline 
			${}^2$H  &    $2.2102$ & $2.2246$ & $2.157$  &  $2.140$		\\
			${}^3$H  &	  $8.35(22)$	& $8.48$ & $1.7641(22)$	   &    $1.759$		\\
			\hline
			${}^3$He &    $7.64(14)$	&    $7.72$ &	$1.9064(84)$	   &   $1.945$			\\
			${}^4$He &    $28.24(16)$	&    $28.3$&  $1.7244(29)$	   &   $1.676$			\\
			${}^6$He &    $29.04(13)$	&    $29.27$&  $2.0399 (228)$	   &   $2.068(11)$			\\
			\hline
			${}^{6}$Li &  $32.82(12)$	&    $31.99$& $2.5484(248)$	   &  $2.589(39)$			\\
			${}^{7}$Li &  $39.61(13)$	&    $39.24$& $2.4496(191)$	   &   $2.444(42)$			\\
			\hline
			${}^{8}$Be &  $56.73(38)$	&    $56.5$& 	   &  			\\
			${}^{9}$Be &  $57.59(29)$	&    $58.17$& $2.5299(337)$	   &  $2.518$			\\
			${}^{10}$Be & $63.72(30)$	&    $64.97$&    &			\\
			\hline
			${}^{10}$B &  $64.46(59)$	&    $64.75$&   2.4852(348)  &  2.4278		\\
			${}^{11}$B &  $75.38(42)$	&    $76.2$&   2.4502(194) &	2.4059		\\
			\hline
			${}^{12}$C$_{0^{+}_{1}}$ &    $92.36(64)$	&    $92.16$ &  $2.4903(121)$ &  $2.470$			\\
			${}^{12}$C$_{0^{+}_{2}}$ &    $84.88(143)$	&    $84.51$ &  &  			\\
			${}^{12}$C$_{2^{+}_{1}}$ &    $87.58(101)$	&    $87.72$ &  & 			\\
			${}^{13}$C &    $97.07(52)$	&    $97.11$&  $2.5206(408)$ &  $2.4614$			\\
			${}^{14}$C &    $104.87(69)$	&    $105.28$&  $2.5647(396)$ &  $2.504$			\\
			\hline
			${}^{14}$N &    $106.25(94)$	&    $104.66$&  $2.5748(216)$ &  $2.5579$			\\
			${}^{15}$N &    $115.29(37)$	&    $115.49$&  $2.6589(471)$ &  $2.6061$			\\
			\hline
			${}^{16}$O &    $129.99(38)$	&    $127.62$& $2.6935(180)$  &  $2.701$			\\
			${}^{17}$O &    $132.47(36)$	&    $131.76$& $2.6866(756)$  &  $2.695$			\\
			${}^{18}$O &    $140.37(45)$	&    $139.81$& $2.7656(498)$  &  $2.775$			\\
			${}^{20}$O &    $151.90(29)$	&    $151.37$&   &				\\
			${}^{22}$O &    $160.7(17)$	&    $162.03$&   &				\\
			${}^{24}$O &    $166.9(12)$	&    $168.38$   &	&				\\
			\hline
			${}^{18}$F &    $135.52(61)$	&    $137.369$  &  & 		\\
			\hline
			${}^{20}$Ne &    $164.6(06)$	&    $160.645$& $ 2.9971(319)$  &  $3.0053$			\\
			\hline
			${}^{24}$Mg &    $196.0(19)$	&    $198.257$& $ 3.0604(269)$  &  $	3.0568$		\\
			\hline
			${}^{28}$Si &    $234.2(25)$	&    $236.537$& $ 3.0972(227)$  &  $	3.1223$		\\
			\hline
			${}^{32}$S &     $~~266.8(24)$	&    $271.78$& $ 3.2450(703)$  &  $	3.2608$		\\
			\hline
			${}^{36}$Ar &    $~~299.8(35)$	&    $306.72$& $ 3.4469(1075)$  &  $	3.3902$		\\
			\hline
			${}^{40}$Ca &    $~~337.7(14)$	&    $342.052$& $ 3.4995(875)$  &  $	3.4764$	\\
			\hline    \hline
		\end{tabular}
		\label{tab:RMS-charge_radii}
	\end{table}

	\begin{table}[ht]
		\centering
		\caption{Results of calculated pure neutron matter energy using various number of neutrons and box sizes.}
		\small
		\begin{tabular}{ |c | c | c || c  | }
			\hline\hline 
			\quad~$A(=N)$~\quad     &
			\hspace{0.5cm}$L$~(fm)\hspace{0.5cm}            &  
			\hspace{0.5cm}$\rho$~(fm$^{-3}$)\hspace{0.5cm}            &  
			\hspace{1cm}$E$ ~ (MeV)\hspace{1cm}            \\
			\hline\hline
			$14$	&	$6.58$	 &  $0.0492$	 &  $102.9(1)$  \\
			$14$	&	$7.89$	 &  $0.0285$	 &  $66.84(92)$  \\
			$14$	&	$9.21$	 &  $0.0179$	 &  $51.41(76)$  \\
			$14$	&	$10.5$	 &  $0.0120$	 &  $41.41(85)$  \\
			$14$	&	$11.8$	 &  $0.0084$	 &  $34.11(65)$  \\
			$14$	&	$13.2$	 &  $0.0062$	 &  $28.71(50)$  \\
			\hline
			$28$	&	$6.58$	 &  $0.0984$	 &  $341.5(99)$  \\
			$28$	&	$7.89$	 &  $0.0570$	 &  $220.8(64)$  \\
			$28$	&	$9.21$	 &  $0.0359$	 &  $169.9(35)$  \\
			$28$	&	$10.5$	 &  $0.0240$	 &  $136.0(24)$  \\
			$28$	&	$11.8$	 &  $0.0169$	 &  $112.8(21)$  \\
			\hline
			$42$	&	$6.58$	 &  $0.1477$	 &  $607.4(70)$  \\
			$42$	&	$7.89$	 &  $0.0855$	 &  $359.9(42)$  \\
			$42$	&	$9.21$	 &  $0.0538$	 &  $267.3(43)$  \\
			$42$	&	$10.5$	 &  $0.0361$	 &  $211.1(10)$  \\
			$42$	&	$11.8$	 &  $0.0253$	 &  $175.7(10)$  \\
			$42$	&	$13.2$	 &  $0.0185$	 &  $149.6(22)$  \\
			\hline
			$66$	&	$6.58$	 &  $0.2320$	 &  $1750(7)$  \\
			$66$	&	$7.89$	 &  $0.1343$	 &  $928.4(17)$  \\
			$66$	&	$9.21$	 &  $0.0846$     &  $613.1(44)$  \\
			$66$	&	$10.5$	 &  $0.0567$	 &  $464.0(66)$  \\
			$66$	&	$11.8$	 &  $0.0398$	 &  $373(19)$  \\
			\hline
			$80$	&	$6.58$	 &  $0.2812$	 &  $2429(10)$  \\
			$80$	&	$7.89$	 &  $0.1628$	 &  $1535(7)$  \\
			$80$	&	$9.21$	 &  $0.1025$	 &  $999(8)$  \\
			\hline \hline
		\end{tabular}
		\label{tab:Energy-neutron-matter}
	\end{table}

	\begin{table}[ht]
		\centering
		\caption{Results of calculated symmetric nuclear matter energy using various number of nucleons.}
		\small
		\begin{tabular}{ |c | c | c || c || c  | }
			\hline\hline 
			\quad~$A(=2N=2Z)$~\quad     &
			\hspace{0.5cm}$L$~(fm)\hspace{0.5cm}            &  
			\hspace{0.5cm}$\rho$~(fm$^{-3}$)\hspace{0.5cm}            &
			$E$ at N3LO [2NFs only] ~ (MeV)         &  
			$E$ at N3LO [2NFs+3NFs] ~ (MeV)            \\
			\hline\hline 
			$~12$   &   $9.21$   &    $0.0154$    &   $-101.4(2)$   &   $-102.4(7)$ \\
			$~16$   &   $9.21$   &    $0.0205$    &   $-137.3(3)$   &   $-164(2)$ \\
			$~24$   &   $9.21$   &    $0.0308$    &      $-219.9(1)$   &   $-281(3)$ \\
			$~36$   &   $9.21$   &    $0.0461$    &      $-328(1)$   &   $-456(2)$ \\
			$~48$   &   $9.21$   &    $0.0615$    &      $-453(1)$   &   $-697(4)$ \\
			$~60$   &   $9.21$   &    $0.0768$    &      $-569(2)$   &   $-851(6)$ \\
			$~72$   &   $9.21$   &    $0.0922$    &      $-699(1)$   &   $-1067(5)$ \\
			$~84$   &   $9.21$   &    $0.1076$    &      $-830(2)$   &   $-1328(5)$ \\
			$~96$   &   $9.21$   &    $0.1230$    &      $-965(3)$   &   $-1642(6)$ \\
			$112$   &   $9.21$   &    $0.1435$    &      $-1090(2)$   &   $-1945(8)$ \\
			$128$   &   $9.21$   &    $0.1640$    &      $-1060(2)$   &   $-2078(7)$ \\
			$144$   &   $9.21$   &    $0.1845$    &      $-963(7)$   &   $-2150(43)$ \\
			$160$   &   $9.21$   &    $0.2050$    &      $-850(39)$   &   $-2437(317)$ \\
			\hline    \hline
		\end{tabular}
		\label{tab:Energy-nuclear-matter}
	\end{table}

	%\subsection{Preamble ({in progress})}\label{S-1.0}
	
	\subsection{Nuclear Lattice Effective Field Theory}
	
	Nuclear lattice effective field theory (NLEFT) \cite{Lee:2008fa,Lahde:2019npb} combines the frameworks of chiral effective field theory ($\chi$EFT) for the forces between nucleons, lattice field theory, and stochastic Monte Carlo algorithms. Each of these forms a cornerstone of the modern approach to strongly interacting many-fermion systems in many fields of study, notably nuclear, particle, condensed matter and atomic physics. By a unique merger of these cornerstones, NLEFT has matured into a leading framework for the investigation of nuclear structure properties. 
	
	Building upon the early developments \cite{Borasoy:2006qn}, NLEFT has extended its reach to light and medium-mass nuclei, mostly of even-even type. This has enabled detailed predictions \cite{Meissner:2014lgi,Meissner:2015lca,Lee:2021wey} of nuclear structure properties, in particular the binding energy for \ce{^3H} and \ce{^4He} at NNLO \cite{Krebs:2008nt}, as well as binding energies for isotopic chains with $Z=1,2,4,6$ and $8$ \cite{Elhatisari:2017eno}, together with root-mean-square radii and proton and neutron densities.
	Moreover, NLEFT has given further impulse to the investigation of $\alpha$-clustering in nuclear matter. The analyses of the Hoyle state \cite{Epelbaum:2012qn,Lahde:2013png} and further low-lying excited states \cite{Lahde:2013png,Shen:2022bak} of \ce{^{12}C}, as well as the study of the structure and EM properties of the $0_1^+$, the $0_2^+$ and $2_1^+$ states of \ce{^{16}O} \cite{Epelbaum:2013paa}, and the prediction of ground-state properties of even-even self-conjugate nuclei with $Z \leq 14$ \cite{Lahde:2013uqa}, are noteworthy in this respect.
	
	An immediate strength of NLEFT is the favorable computational scaling with nucleon number~$A$, as the $A$-body Hamiltonian in NLEFT is diagonalized stochastically by means of an auxiliary field Monte Carlo (AFMC) algorithm. The two-body, three-body, and pion exchange interactions are described by auxiliary fields, which are sampled at each step in the AFMC algorithm. Energy eigenvalues are obtained using the adiabatic projection method, whereby transition amplitudes mediated by the Schr\"odinger time evolution operator between trial (Slater-determinant) states are constructed. In practical AFMC calculations of NLEFT, the temporal evolution of the trial states is discretized into $L_t$ equal steps, separated by a temporal lattice spacing $a_t$. Hence, the time-evolution operator is subdivided into a sequence of $L_t$ transfer matrices (Trotter decomposition). 
	
	On the one hand, NLEFT is firmly grounded in low-energy QCD, as the interactions between nucleons are described by the NNLO or N3LO Lagrangians of $\chi$EFT. The resulting Hamiltonian is formulated with a finite cubic lattice regulator, and avoids the need for \textit{pre-diagonalization} techniques such as the similarity renormalization group (SRG) \cite{Tichai:2020dna}. On the other hand, it is known that the AFMC implementation of the full NNLO or N3LO $\chi$EFT Hamiltonian produces a severe sign problem, which eventually compromises the effectiveness of the numerical technique. 
	
	A firmly-rooted remedy in the NLEFT literature \cite{Lee:2008fa} entails the replacement of the $\chi$EFT transfer matrices in the initial and the final $L_{t_o}$ time steps by pionless $SU(4)$-symmetric transfer matrices. It follows that the full $\chi$EFT Hamiltonian acts only in the middle $L_{t_i} \equiv L_{t}-L_{t_o}$ time steps through the transfer matrices, whereas Wigner's SU(4) one acts as a low-energy filter \cite{Lee:2008fa} at the boundaries of the time interval.  This procedure extends to the expectation value of any operator representing a physical observable, inserted in the midpoint of the chain of transfer-matrices. The positiveness of the transfer matrices, where Wigner's SU(4) action is used, is the major advantage of the method, and is guaranteed for systems with even number of nucleons and either spin-singlet or isospin-singlet quantum numbers \cite{Lee:2008fa}. However, distortions in the energy eigenvalues as a result of the usage of an isospin-preserving SU(4) Lagrangian in the action do appear in the form of lower bounds \cite{Lee:2008fa}. 
	Although remedies such as eigenvector continuation \cite{Frame:2017fah} or symmetry-sign extrapolation \cite{Lahde:2015ona} would allow for a broader use of the $\chi$EFT transfer matrix, in this work we tackle the problem by using the method of wave function matching.

	\subsection*{Other \textit{ab initio} methods}
	Over the past two decades, nuclear \textit{ab initio} methods have had great success. 
	The widespread adoption of nuclear forces from Effective Field Theory (EFT)~\cite{Weinberg:1990,Weinberg:1991, Kolck:1994, Epelbaum:2008ga, Machleidt:2011, Entem:2017, Hammer:2020, Epelbaum:2020} laid the foundation for these developments. More specifically, the inclusion of chiral two-nucleon to higher orders~\cite{Epelbaum:2008ga, Machleidt:2011, Entem:2017, Ekstrom:2015} and the leading three-nucleon~\cite{Kolck:1994, Navratil:2007, Roth:2011, Hebeler:2020ocj} forces have shown to be key elements for high quality calculations.
	The strong short-range repulsion in nuclear potentials is often soften by a renormalization procedure~\cite{Bogner:2010}. Besides Nuclear Lattice Effective Field Theory (NLEFT),  there are many other promising \textit{ab initio} approaches being used to calculate the properties of few- and many- nucleon systems.
	Full configuration methods like no-core shell model (NCSM)~\cite{Bruce:2013, Roth:2014, Jurgenson:2013, LENPIC:2022cyu}, symmetry-adapted NCSM \cite{Dytrych:2020,Dreyfuss:2020} and quantum Monte Carlo (QMC) in several different varieties~\cite{Carlson:2015, Pastore:2017uwc, Lynn2019, Gandolfi:2020pbj, Schiavilla:2021dun} were able to extend their reach into the lower $sd$-shell.
	
	With controlled truncations, a variety of computationally efficient techniques were developed, like self-consistent Green's Functions (SCGF)~\cite{Soma:2020xhv}, closed-shell many-body perturbation theory (MBPT)~\cite{Roth:2010, Tichai:2016}, coupled cluster (CC)~\cite{Hagen:2013nca}, in-medium SRG (IMSRG)~\cite{Hergert:2016}, and {\it ab initio} no-core Monte Carlo shell model \cite{Abe:2021sky,Otsuka:2022bcf}.
	In the meantime, effective shell model Hamiltonians can be constructed by open-shell MBPT~\cite{Morten:1995, Coraggio:2009, Tichai:2018}, valence-space IMSRG (VS-IMSRG) \cite{Bogner:2014, Stroberg:2016ung} and shell model coupled cluster (SMCC) \cite{Jansen:2014,Sun2018}.
	To include continuum effects in weakly bound nuclei, several methods were also well established, like the no-core shell model with continuum (NCSMC)~\cite{Navratil:2016}, complex CC~\cite{Hagen2012}, no-core Gamow shell model (GSM)~\cite{Papadimitriou:2013, Li:2019}, Gamow IMSRG~\cite{Hu:2019} and GSM with realistic forces \cite{Sun:2017,Ma:2020}. It should be noted that the level of the approximation is not the same in all these different approaches, but a detailed comparison goes beyond the scope of this paper.

	\bibliography{References}
	\bibliographystyle{apsrev}
	
\end{document}